\documentclass[useAMS,usenatbib]{mn2e}

\usepackage{graphics}
\usepackage{epsfig}
\usepackage{latexsym}

\usepackage{amsfonts}
\usepackage{amsmath}
\usepackage{amssymb}
\usepackage[english]{babel}
\usepackage{graphicx}
\usepackage{url}
\usepackage{textcomp}

\usepackage{amssymb}
\usepackage{graphics}
\usepackage{graphicx}
\usepackage{epstopdf}

\usepackage[usenames]{color}

\newcommand{\be}{\begin{equation}}
\newcommand{\ee}{\end{equation}}
\newcommand{\ba}{\begin{eqnarray}}
\newcommand{\ea}{\end{eqnarray}}
\newcommand{\bc}{\begin{center}}
\newcommand{\ec}{\end{center}}

\title[Time-dependent modeling of pulsar wind nebulae]{Time-dependent modeling of pulsar wind nebulae:\\
Study on the impact of the diffusion-loss approximations}
\author[Mart\'in, Torres, Rea]{Jonatan Mart\'in$^{1}$, Diego F. Torres$^{1,2}$, \& Nanda Rea$^{1}$\\
$^1$Institut de Ci\`encies de l'Espai (IEEC-CSIC), Campus UAB,  Torre C5, 2a planta, 08193 Barcelona, Spain \\
$^2$Instituci\'o Catalana de Recerca i Estudis Avan\c{c}ats (ICREA) Barcelona, Spain}

\begin{document}

\date{}

\pagerange{\pageref{firstpage}--\pageref{lastpage}} 
\pubyear{2012}

\maketitle

\label{firstpage}

\begin{abstract}
In this work, we present a
leptonic, time-dependent model of pulsar wind nebulae (PWNe).
The model seeks a 
solution for the lepton distribution function
considering the full time-energy dependent diffusion-loss equation.
The time-dependent lepton population is balanced by injection, energy losses, and escape. 
We include synchrotron, inverse Compton (IC, with the cosmic-microwave background as well as with IR/optical photon fields), 
self-synchrotron Compton (SSC), and
bremsstrahlung processes, all devoid of any radiative approximations. 
With this model in place we focus on the Crab nebula as an example and present its time dependent evolution. Afterwards, we analyze the impact of
different approximations made at the level of the diffusion-loss equation, as can be found in the literature. 
Whereas previous models  ignored the escape term, e.g., with the diffusion-loss equation becoming advective, others approximated the losses as catastrophic, so that the equation has only time derivatives. 
Additional approximations are also described and computed.
We show which is the impact of these approaches in the determination of the PWN evolution. In particular, we find the time-dependent deviation
of the multi-wavelength spectrum and the best-fit parameters obtained with the complete and the  approximate  models.  
\end{abstract}


\section{Introduction}

In addition to their electromagnetic emission, pulsars dissipate their
rotational energy via relativistic winds of particles. Because the
relativistic bulk velocity of the wind is supersonic with respect to the
ambient medium, such a wind produces a termination shock. In turn, the wind
particles, moving trough the magnetic field and the ambient photons,
produce radiation that we observe as Pulsar Wind Nebulae (PWNe). As
the pulsars themselves, the PWN emits at all wavelengths from radio to
TeV.

PWNe usually have two main X-ray morphologies, depending to the
velocity of the pulsar proper motion. For slow pulsars, images taken with the {\em Chandra} X-ray Observatory (see e.g., Kargaltsev \& Pavlov, 2008) show a toroidal shape around
the pulsar equator, with two possible jets starting from the pulsars
poles. Instead, pulsars moving with high velocity in the interstellar
medium produce PWNe with the characteristic bullet-like or bow-shock
morphology, with the tail developed along the pulsar motion. Thus, the study
of PWNe can lead to knowledge of pulsar winds, the properties of the
ambient medium, and the wind-medium interaction.

PWNe constitute the largest class of identified Galactic very-high energy (VHE)
gamma-ray sources, with the number of TeV detected objects increasing from 1 to
$\sim$30 in the last 6 years. This statistics shines in comparison
with the $\sim$30, 10, or 40 PWNe known in radio, optical/IR, or
X-rays, respectively, detected in decades of observations. The
majority of PWNe at VHE gamma-rays have a very large size, depending on
their evolutionary stage and proximity (see e.g., Vela X or HESS
J1825-137). The Cherenkov Telescope Array (CTA, Actis et al. 2012),
with its large FoV, will be able to image the whole plerionic emission,
including the halo produced by cooled electrons. The larger energy
range of CTA compared with all other gamma-ray instruments will
be crucial to understand cooling effects, including the resolution of
internal structures and the properties which could be therein active,
such as their likely enhanced magnetic field, and the possibility of
disentangling between synchrotron and adiabatic losses. For the first
time, we will be able to study in detail the interaction between the
host remnant and its PWN, and how much of the gamma-ray emission
originates in each component should feedback the theoretical
understanding of the evolutionary tracks. Observations with CTA will
hopefully produce a homogeneous sampling of the Galactic PWNe, since its
sensitivity will permit the detection of PWNe up to 50 kpc. Out of the
complete PWNe population in the Galaxy, assuming 40 kyr for the
estimated lifetime of TeV-emitting electrons for a magnetic field of 3
micro Gauss, CTA would detect 300--600 objects, of which between
15-25\% will be fully resolved, depending on proximity, age, and flux
(see  de O\~na et al. 2012 for detailed studies on CTA expectations).

In studying PWNe, there are two distinct theoretical approaches. On
the one hand, detailed magneto-hydrodynamic (MHD) simulations have
succeeded in explaining the morphology of PWNe (see Rea \& Torres 2011
for reviews on several aspects of these issues). On the other hand, spherically symmetric 1D PWNe spectral
models, with no energy-dependent morphological output, have been
constructed since decades (although there are
only a handful of such codes, and none is public to our knowledge),
see, e.g., Aharonian et al. 1997, Atoyan \& Aharonian 1998, Bednarek \& Bartosik
2003, ibid. 2005;
B\"usching
et al. 2008; Fang \& Zhang 2010; Zhang et al. 2008; Tanaka \& Takahara
2010, 2011;  Li et al. 2010; and other references quoted below. 

Some of the most elaborate models solve the time-energy dependent diffusion-loss
equation, where the time-dependent lepton population is balanced by
injection, energy losses, and escape, with various degrees of
approximations. Such models of PWNe  usually calibrate with the Crab nebula
measurements today, being it the best studied PWN at all wavelengths. 
Whereas essentially all models properly fit the Crab
nebula data at the current age, 
the time-evolution out of the normalization age of the PWN  can show significant deviations. 
In this work
we present a
leptonic, time-dependent model of pulsar wind nebulae (PWNe). It seeks
a solution for the electron distribution function
considering the full time-energy dependent diffusion-loss equation.
The time-dependent lepton population is balanced in our model by
injection, energy losses, and escape.
We include synchrotron, Inverse Compton (with the cosmic-microwave
background as well as IR/optical photon fields),
self-synchrotron Compton, and
bremsstrahlung processes, all devoid of any radiative approximations.
With this model in place we focus on the Crab nebula as an example and
achieve a fit for its persistent emission.
Afterwards,
we analyze the impact of
different approximations made at the level of the diffusion-loss
equation, as can be found in the literature.
We note that conclusions
on specific sources and population analysis using  approximate  tools
is severely affected.  In \S 2 we describe our model. In \S 3,
we show the results obtained for the Crab nebula and the luminosity ratios
computed by our code and compared with observations. In \S 4, we comment on 
different approximations found in the literature and
we show their impact in the Crab nebula evolution. Finally, in \S 5, we
write our concluding remarks.

\section{Modeling the  emission of PWNe}

\subsection{The diffusion equation}
\label{diffusion}

The diffusion equation adopted in this work reads (e.g., Ginzburg \& Syrovatsky 1964)
\begin{equation}
\label{te}
\frac{\partial N(\gamma,t)}{\partial t}=-\frac{\partial}{\partial \gamma}\left[\dot{\gamma}(\gamma,t)N(\gamma,t) \right]-\frac{N(\gamma,t)}{\tau(\gamma,t)}+Q(\gamma,t),
\end{equation}
where 
the left-hand side is the variation of the lepton distribution in time and
the first term in the right-hand side accounts for the continuous change in energy of the particles due to energy losses. The function $\dot{\gamma}(\gamma,t)$ is the summation
of the energy losses due to all processes considered. $Q(\gamma,t)$ represents the injection of particles
per unit energy and unit volume in a certain time.
$\tau(\gamma, t)$ represents the escape time, 
after which the particles are effectively removed from the phase space.
We assume that particles escape (second term in the right hand side of Eq. 1) via Bohm diffusion (e.g., as in 
Zhang et al. 2008, or Li et al. 2010).
Eq. (\ref{te}) is solved using the implicit forward difference technique on
the derivatives in time and energy. 

\subsection{The injection of particles}

Our numerical implementation allows for any form of particle injection.
We assume a broken power-law
\begin{equation}
\label{injection}
Q(\gamma,t)=Q_0(t)\left \{
\begin{array}{ll}
\left(\frac{\gamma}{\gamma_b} \right)^{-\alpha_1}  & \text{for }\gamma \le \gamma_b,\\
 \left(\frac{\gamma}{\gamma_b} \right)^{-\alpha_2} & \text{for }\gamma > \gamma_b,
\end{array}  \right .
\end{equation}
where $\gamma_b$ is the break energy. The parameters $\alpha_1$  and $\alpha_2$
are the spectral indices. The normalization constant $Q_0(t)$ is determined using the injection luminosity $L(t)$, 
\begin{equation}
\label{injectionevol}
L(t)=L_0 \left(1+\frac{t}{\tau_0} \right)^{-\frac{n+1}{n-1}},
\end{equation}
and where $L_0$ is the initial luminosity, $\tau_0$ is the initial spin-down timescale of the pulsar and $n$ is the breaking index. These parameters may be observationally determined (see, e.g., Lewin  \& van der Klis  2006, Gaensler \& Slane 2006). In the case of the spin-down luminosity, we have
\begin{equation}
\label{edot}
L(t)=4\pi^2 I \frac{\dot{P}}{P^3},
\end{equation}
where $P$ and $\dot{P}$ are the period and its first derivative and $I$ is the pulsar
moment of inertia, which we assume $I \sim 10^{45}$ g cm$^2$. 
The initial spin-down timescale of the pulsar is (Gaensler \& Slane 2006)
\begin{equation}
\label{spindownage}
\tau_0=\frac{P_0}{(n-1)\dot{P}_0}=\frac{2\tau_c}{n-1}-t_{age},
\end{equation}
where
$P_0$ and $\dot{P}_0$ are the initial period and its first derivative and $\tau_c$ is the characteristic age of the pulsar, 
\begin{equation}
\label{chaage}
\tau_c=\frac{P}{2\dot{P}}.
\end{equation}
The breaking index can also be computed from observational data when the second derivative of the period, $\ddot{P}$, is known.
Assuming that the angular frequency $\Omega=2\pi/P$ of the pulsar evolves in time as
$
\dot{\Omega}=k \Omega^n
$
where $n$ is again the breaking index and $k$ is a constant that depends on the magnetic moment of the pulsar, we find
$
n={\Omega \ddot{\Omega}}/{\dot{\Omega}^2} \simeq {P \ddot{P}}/{\dot{P}^2}.
$
If the system is a dipole spin-down rotator, the breaking index is exactly 3 and the constant $k$ has the value
$
k={2\mu_\perp^2} / {3Ic^3},
$
where $\mu_\perp$ is the component of the magnetic dipole moment orthogonal to the rotation axis.

The normalization of the injection function is given by
\begin{equation}
\label{normalization}
(1-\eta)L(t)=\int_0^\infty \gamma m c^2 Q(\gamma,t) \mathrm{d}\gamma,
\end{equation}
where $\eta$ is the magnetic energy fraction. It is defined as $\eta=L_B(t)/L(t)$, where $L_B(t)$ is the magnetic power; thus, $\eta$ is its ratio with the spin-down power. This definition, see e.g., Tanaka \& Takahara (2010), divides the energy injection from the pulsar into magnetic field energy and relativistic particle energy and is different from the one used for 
the magnetization parameter 
$
\sigma(t)= {L_B(t)} / {L_p(t)},
$
and where $L_p(t)$ is the relativistic particle's fraction of the spin-down power. 
To ensure particle confinement, we impose that the Larmor radius of the particles has to be smaller than the termination shock radius, what leads to \begin{equation}
\gamma_{max}(t)=\frac{\varepsilon e \kappa}{m_e c^2}\sqrt{\eta \frac{L(t)}{c}},
\end{equation}
where $e$ is the electron charge and $\varepsilon$ is the fractional size of the radius of the shock, which has to be smaller than 1. $\kappa$ is the magnetic compression
ratio. For strong shocks ($\sigma \ll 1$), $\kappa \simeq 3$ (Venter \& de Jager 2006, 
Holler et al. 2012). 

Simulations with relativistic shocks in unmagnetized plasmas predict a particle spectrum downstream that has two components: a relativistic Maxwellian and a high-energy tail fitted by a power-law with an energy index of $-2.4 \pm 0.1$ (Spitkovsky 2008). Some papers extrapolate these results to the case of PWNe, see, for instance, Grondin et al. 2011 and Van Etten \& Romani 2011. However, Spitkovsky's simulations of relativistic shocks in unmagnetized plasmas accelerate particles until $\gamma \approx 1000$ only. 
Here, for simplicity and to facilitate comparison with other models (see e.g., Zhang et al. 2008, Gelfand et al. 2009 or Tanaka \& Takahara 2010), we consider a pure broken power-law injection shown in Eq.~(\ref{injection}).

\subsection{Energy losses and photon spectrum}

We consider  
synchrotron, (Klein-Nishina) inverse Compton, bremsstrahlung, and adiabatic losses (and their time dependence).
Details of their implementation are given in the Appendix.

\section{Results}

\subsection{The Crab nebula}

The distance to the Crab Nebula  is 2 kpc (Manchester et al. 2005). The period and
 its derivative are obtained from Taylor et al. (1993). 
  Assuming that the moment of inertia of the Crab pulsar is $I=10^{45}$ g cm$^2$, and using
 Eq. (\ref{edot}), we obtain the spin-down luminosity power today.
The expansion of the PWN is considered using the free expansion approximation given by van
der Swaluw (2001). We consider a characteristic energy
for the SN explosion of $10^{51}$ erg and an ejected mass of 9.5$M_\odot$ (Bucciantini et al. 2011).
All this parameters used for the Crab Nebula are summarized in Table~\ref{crabconstrains} including those
determined by the code.
 
 At the date of the pulsar period ephemeridis (year 1994), the age of the pulsar was 940 yr.
 This is consistent with Eq. (\ref{edot}) and helps to minimize the bias produced by the non-simultaneity of the multi-wavelength  data points used, obtained from $\sim$1970 (radio) to
2008 (VHE). We checked that changing the ephemeris to the latest one (e.g., the one used by the Fermi-LAT Collaboration\footnote{ \url {http://fermi.gsfc.nasa.gov/ssc/data/access/lat/ephems/}})
introduce no visible change in the results.

\begin{table*}
\centering
\scriptsize
\caption{Summary of the physical magnitudes used or obtained for the Crab Nebula fit at the current age. A few parameters are
fixed based on prior input or hypothesis. }
\vspace{0.2cm}
\label{crabconstrains}
\begin{tabular}{llll}
\hline
Magnitude & Symbol & Value & Origin or Result\\
\hline
Age (yr) & $t_{age}$ & 940 & fixed\\
Period (ms) & $P(t_{age})$ & 33.4033474094 & from Taylor et al. (1993)\\
Period derivative (s  s$^{-1}$) & $\dot{P}(t_{age})$ & 4.209599 $\times 10^{-13}$ & from Taylor et al. (1993)\\
Spin-down luminosity now (erg/s) & $L(t_{age})$ & $4.53 \times 10^{38}$ & Eq. (\ref{injectionevol})\\
Moment of inertia (g cm$^2$) & $I$ & $10^{45}$ & Eq. (\ref{edot})\\
Breaking index & $n$ & 2.509 & from Lyne et al. (1988)\\
Distance (kpc) & $d$ & 2 & from Manchester et al. (2005)\\
Ejected mass ($M_\odot$) & $M_{ej}$ & 9.5 & from Bucciantini et al. (2011)\\
SN explosion energy (erg) & $E_0$ & $10^{51}$ & from Bucciantini et al. (2011)\\
\hline
Minimum energy at injection & $\gamma_{min}$ & $1$ & assumed \\
Maximum energy at injection at $t_{age}$ &  $\gamma_{max}(t_{age})$ & $7.9 \times 10^9$ & result of the fit\\
Break energy & $\gamma_b$ & $7 \times 10^5$ & result of the fit\\
Low energy index & $\alpha_1$ & 1.5 & result of the fit\\
High energy index & $\alpha_2$ & 2.5 & result of the fit\\
Shock radius fraction & $\varepsilon$ & $1/3$ & result of the fit\\
\hline
Initial spin-down luminosity (erg/s) & $L_0$ & $3.1 \times 10^{39}$ & result of the fit\\
Initial spin-down age (yr) & $\tau_0$ & 730 & Eq. (\ref{spindownage})\\
Magnetic field ($\mu$G) & $B(t_{age})$ & 97 & result of the fit\\
Magnetic fraction & $\eta$ & 0.012 & result of the fit\\
PWN radius today (pc) & $R_{PWN}(t_{age})$ & 2.1 & Eq. (\ref{rpwn})\\
\hline
CMB temperature (K) & $T_{CMB}$ & 2.73 & fixed\\
CMB energy density (eV/cm$^3$) & $w_{CMB}$ & 0.25 & fixed \\
FIR temperature (K) & $T_{FIR}$ & 70 & as in Marsden et al. (1984) and subsequent refs.\\
FIR energy density (eV/cm$^3$) & $w_{FIR}$ & 0.5 & as in Marsden et al. (1984)  and subsequent refs.\\
NIR temperature (K) & $T_{NIR}$ & 5000 & as in Aharonian et al. (1997) and subsequent refs.\\
NIR energy density (eV/cm$^3$) & $w_{NIR}$ & 1 & as in Aharonian et al. (1997) and subsequent refs.\\
Hydrogen density (cm$^{-3}$) & $n_H$ & 1 & assumed\\
\hline
\hline
\end{tabular}
\end{table*}

\begin{figure*}
\begin{center}
\includegraphics[scale=0.45]{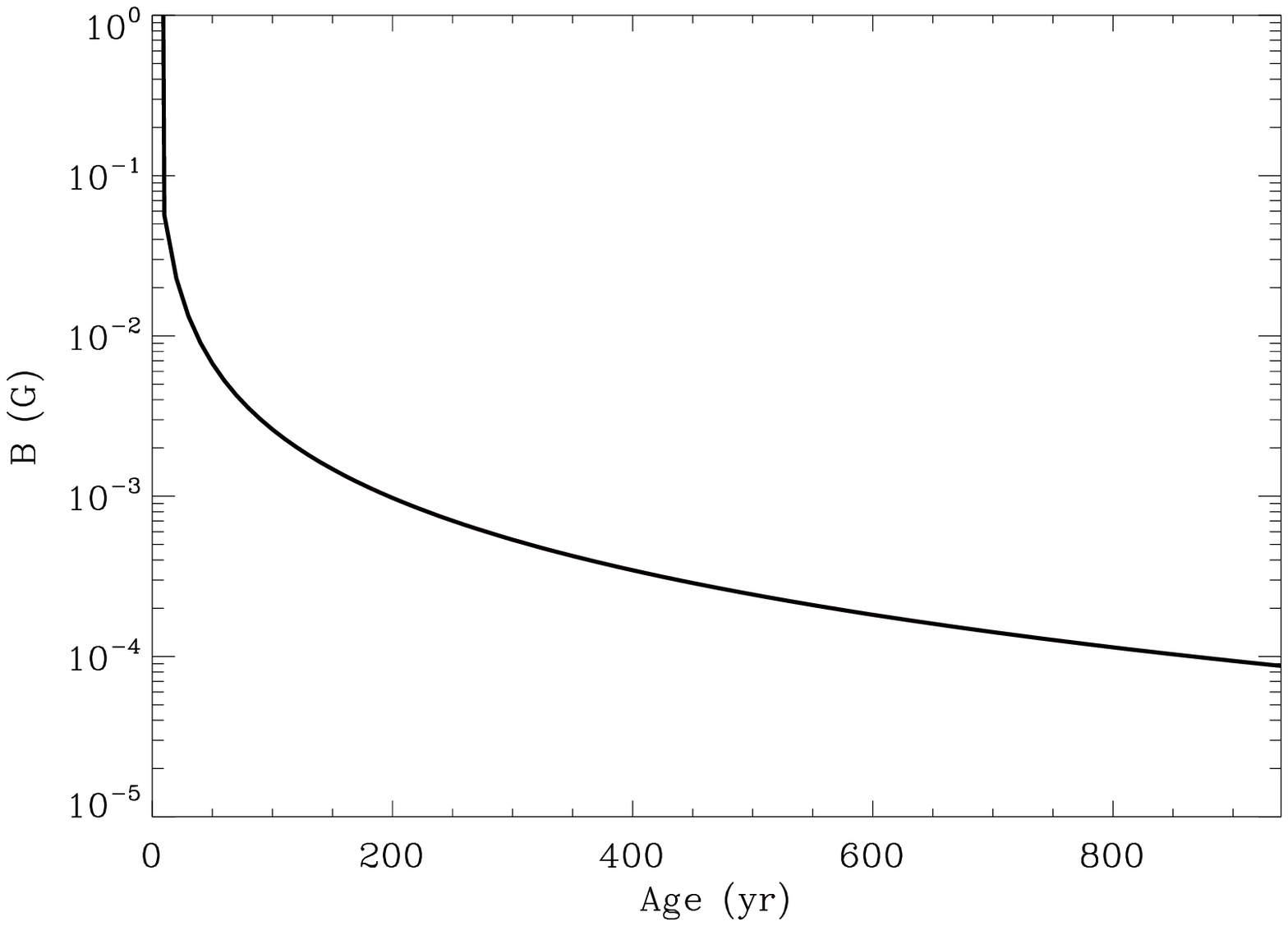}
\includegraphics[scale=0.45]{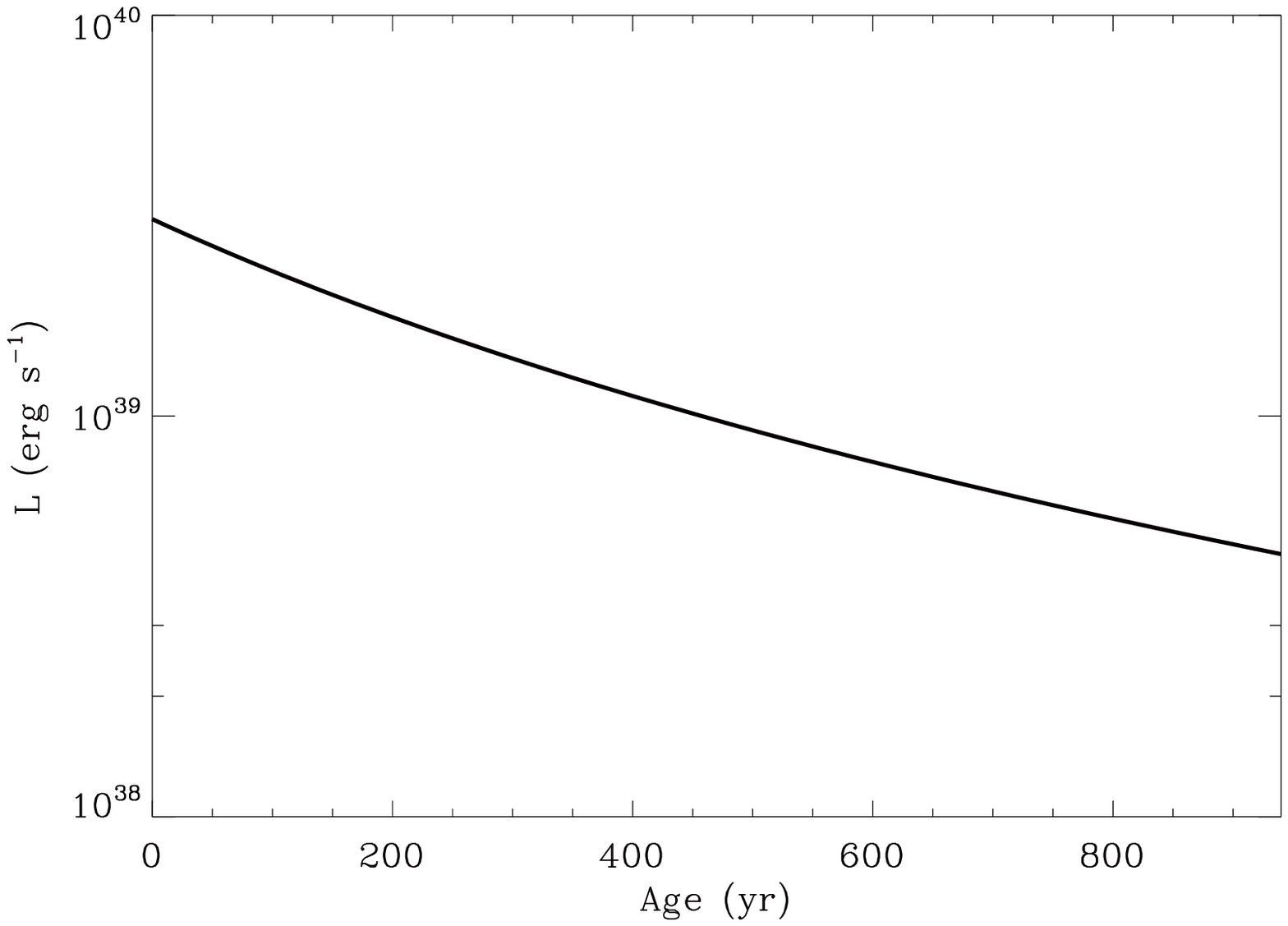}
\includegraphics[scale=0.45]{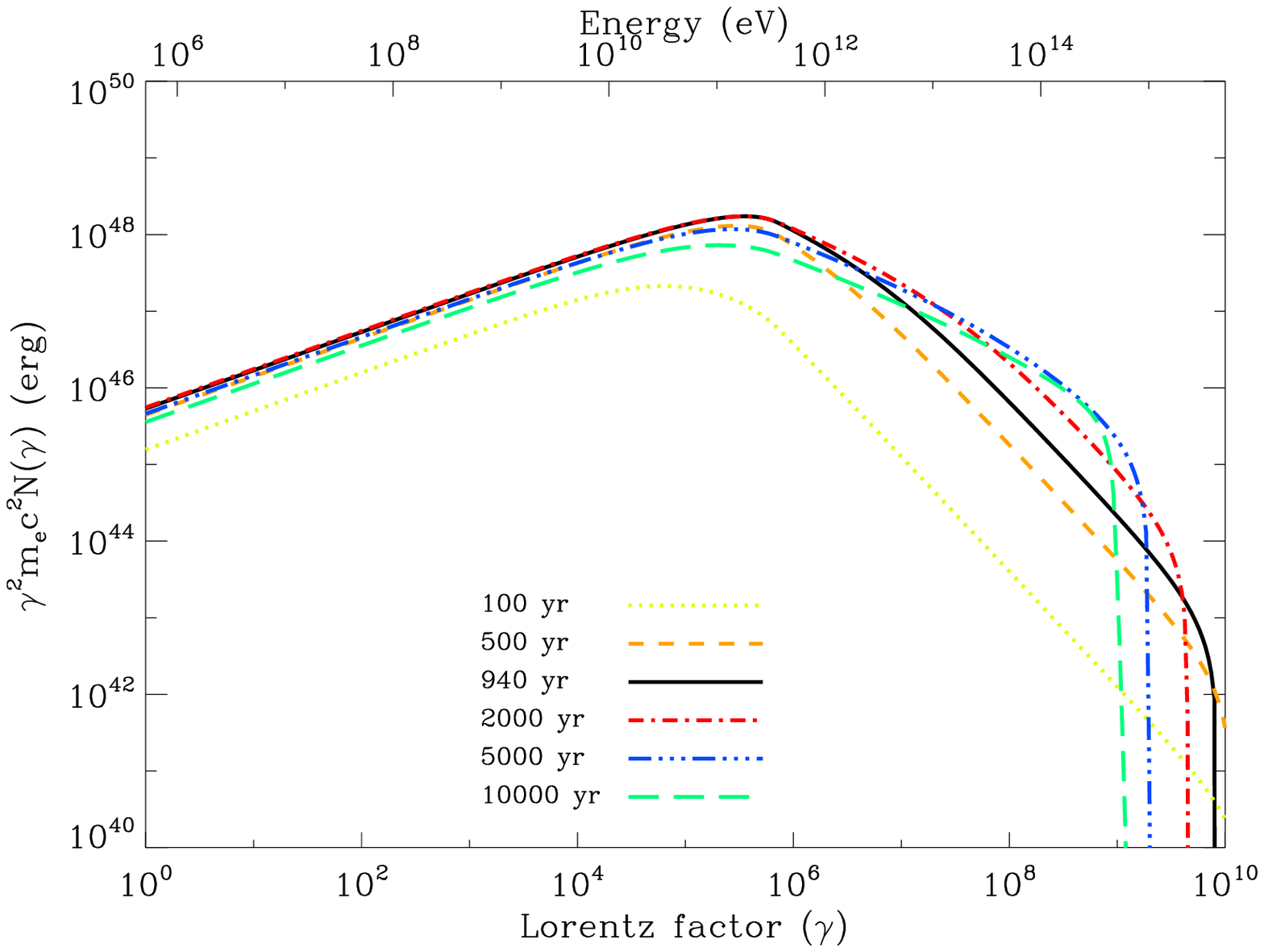}
\includegraphics[scale=0.45]{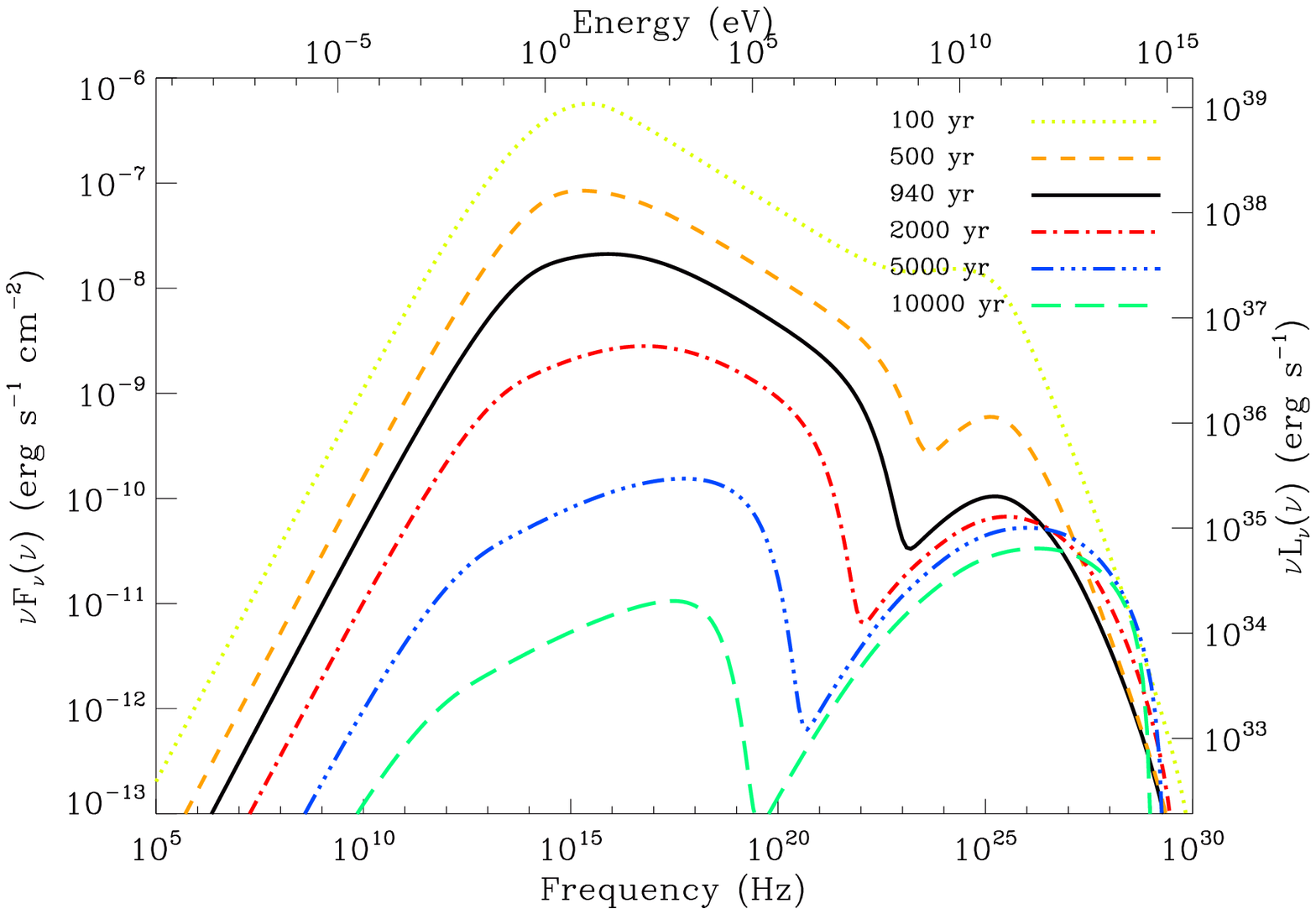}
\end{center}
\caption{From top to bottom, left to right: Magnetic field, spin-down power, lepton population, and spectral energy distribution 
of the Crab nebula as a function of time.}
\label{crab-time}
\end{figure*}

In order to compute the IC energy losses and spectrum, we consider three components: the cosmic microwave background (CMB), the galactic far infrared background (FIR) and the
 near infrared and optical photon field due to the stars (NIR),
 $
 n(\nu)=n_{CMB}(\nu)+n_{FIR}(\nu)+n_{NIR}(\nu).
$
Each one of the latter two is considered as a diluted blackbody (Schlickeiser 2002). 
The temperature of the FIR (NIR) is considered as 70 (5000) K. The CMB is  a blackbody of temperature 2.73 K.

We find different ways to evolve the magnetic field in the literature (see Rees \& Gunn 1974, Kennel \& Coroniti 1984, Venter et al. 2006, de Jager et al. 2009, Tanaka \&
Takahara 2010,2011). We assume magnetic energy conservation as in Tanaka \& Takahara 2010,
\begin{equation}
\int_0^t \eta L(t') \mathrm{d}t'=\frac{4 \pi}{3}R_{PWN}^3(t) \frac{B^2(t)}{8 \pi},
\end{equation}
thus, using Eq. (\ref{injectionevol}) and solving for the field we obtain
\begin{equation}
\label{magevol}
B(t)=\sqrt{\frac{3(n-1)\eta L_0 \tau_0}{R^3_{PWN}(t)}\left[1-\left(1+\frac{t}{\tau_0} \right)^{-\frac{2}{n-1}} \right]}.
\end{equation}
The magnetic energy is conserved in the sense that the magnetic fraction (the fraction of the spin down power that goes into the magnetic field) is constant. This parameter is called $\eta$ in Eq. (10). The magnetic field itself is thus time dependent, and its behavior is given in Eq. (\ref{magevol}). This approach has a very similar behaviour as it is adopted in other papers cited before.

For the injection we use Eq. (\ref{injection}),
where $Q_0(t)$ is calculated using Eq. (\ref{normalization}). The final luminosity power is given by Eq. (\ref{edot}) and the initial spin-down power is determined using Eq. (\ref{injectionevol}), since we know the
luminosity power nowadays and the age of the pulsar. For the ISM density in the Crab Nebula, we take a fiducial value of 1 cm$^{-3}$. 
Thus, our free parameters in order to fit the spectrum are the magnetic fraction $\eta$ and the shock radius fraction $\varepsilon$.
For the final fit, we get $\eta=0.012$ and $\varepsilon=1/3$. The magnetic field value we get today is 97 $\mu$G, which is close to the 100 $\mu$G calculated by the
MHD simulations done by Volpi et al. (2008).
Table \ref{crabconstrains} clarifies which parameters comes from observations or assumptions, and which parameters are used to fit the data.

\begin{figure}
\begin{center}
\includegraphics[scale=0.45]{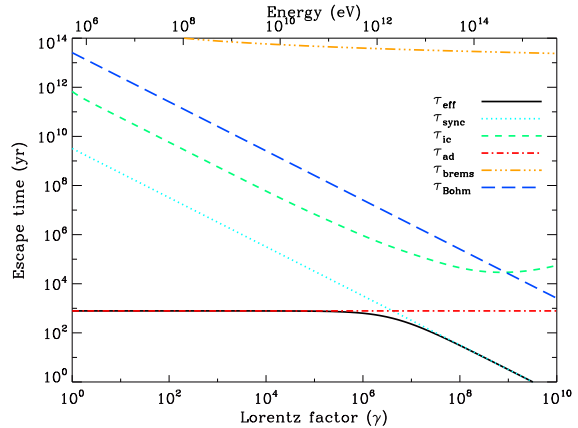}
\end{center}
\caption{Cooling times at $t_{age}=940$ yr. At low energies, the adiabatic losses are dominant because their cooling time is of the same order of the pulsar age. At high
 energies, synchrotron losses become the most important. }
\label{today}
\end{figure}

\subsection{The Crab fitting}
 
\begin{figure*}
\begin{center}
\includegraphics[scale=0.8]{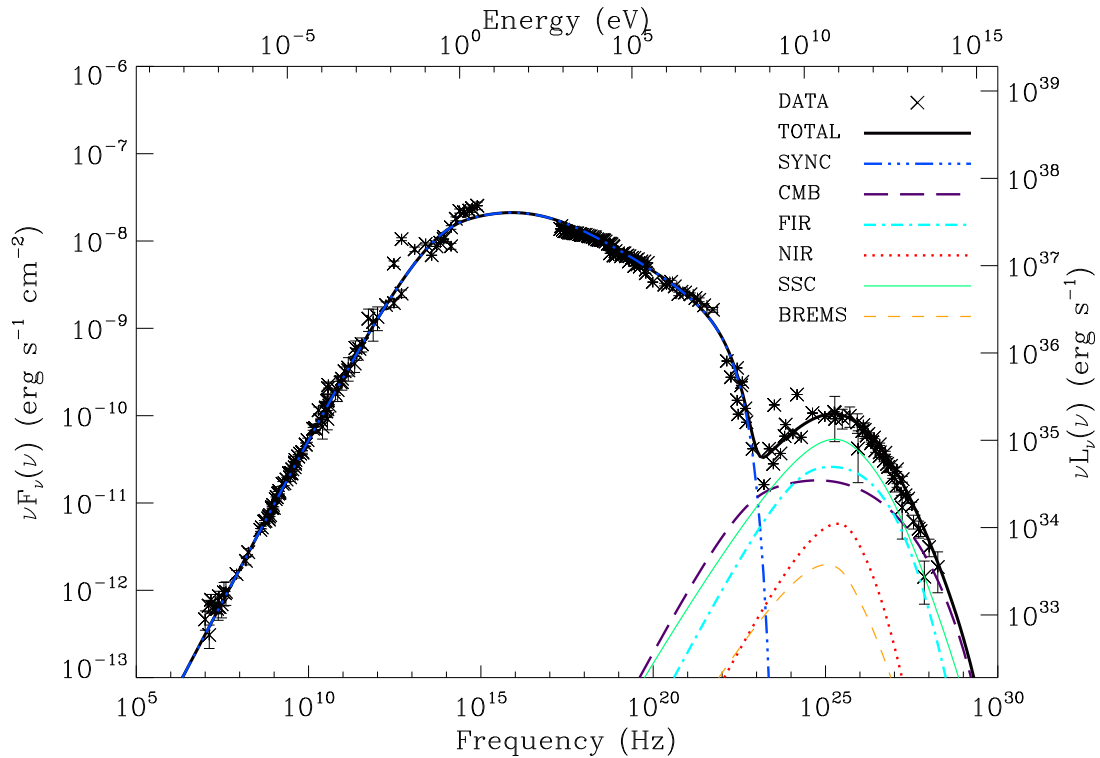}
\end{center}
\caption{Spectrum of the Crab Nebula fitted by our model. The data points are obtained from Baldwin (1971) and Mac\'ias-P\'erez et al. (2010) for
the radio band;
Ney \& Stein (1968), Grasdalen (1979), Geen et al. (2004) and Temim et al. (2006) for the infrared;
Veron-Cetty \& Woltjer (1993) for the optical;
Hennessy (1992) for the ultraviolet,
Kuiper (2001) for the X-rays and soft $\gamma$-rays;  and
Abdo et al. (2010), Aharonian et al. (2004), Aharonian et al. (2006), and Albert et al. (2008) and for the gamma-rays.}
\label{today2}
\end{figure*}

Figure~\ref{crab-time} shows the magnetic field, spin-down power, lepton population, and spectral energy distribution 
of the Crab nebula as a function of time, resulting from our code after normalization to current measurements. The current cooling times for the different processes considered are shown in Figure \ref{today}, whereas the current spectrum is shown 
in Figure~\ref{today2}.

The SSC flux is the strongest contributor to the high-energy spectra, followed by IC with the CMB and the FIR. The Bremsstrahlung contribution is not
very important, but as it is similar to the NIR radiation, we do not neglect it in favor of the other contributions.  Most of the radiative considerations of Tanaka  \& Takahara (2010) are similarly obtained in our model,
since they are driven by SSC domination. 
Our resulting value of the magnetic field today is lower than that used by Atoyan \& Aharonian (1996) in their time-independent approach, who in turn adopted it from the Kennel \& Coroniti (1984) model, followed by an adjustment on the relativistic particle density to enable the data fitting. This value of magnetic field is unrealistically high for our time-dependent spectral model, and a lower value is preferred also by MHD simulations.

\begin{figure}
\begin{center}
\includegraphics[scale=0.45]{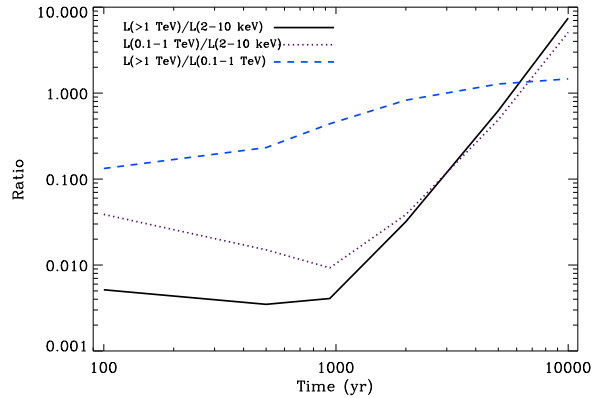}
\end{center}
\caption{Luminosity ratios for the Crab nebula evolution: $L(>$1 TeV) / $L$(2 -- 10 KeV), $L$(0.1
-- 1TeV) / $L$(2 -- 10 KeV) and $L(>$1 TeV) / $L$(0.1 -- 1 KeV)}
\label{ratiolum}
\end{figure}

Regarding the time evolution presented in Figure \ref{crab-time}, it is interesting to note how the peak of the electron distribution moves from lower Lorentz factors to
the energy break in the injection. This displacement of the peak is due to the high energy losses for energies lower than the break at early ages. The maximum energy of
the injection is decreasing with time and the maximum energy of the electrons population is decreasing  also through energy losses, but at a slower rate due to the presence of
high-energy electrons that were previously injected. The slope of the distribution at VHE becomes flattened with time also due to evolution in time of the dominant cooling process,
increasing the power of the IC radiation.
As the magnetic field falls, the synchrotron radiation diminishes with respect the IC radiation and at later ages (e.g., towards 10 kyr), the IC radiation
contains most of the emitted flux. This is in agreement with the idea of older PWNe being still detectable at high energies but being devoid of lower-energy counterparts (de Jager \& Djannati-Ata\"i 2008).
This is shown in Figure \ref{ratiolum}. We see that the flux at
energies $>$1 TeV and the gamma-ray flux are equal for an age of $\sim$5 kyr.

The radio and optical evolution of the Crab nebula show a decreasing-with-time behavior.
Measurements of the radio flux decrease were done by Vinyaikin (2007), using data from
1977 to 2000 at 86, 151.5, 927 and 8000 MHz. The mean flux-decrease rate averaged obtained
was -0.17$\pm$0.02\% yr$^{-1}$. Using data obtained from our code at the same frequencies for the
same time interval we obtained an averaged rate of -0.2\% yr$^{-1}$. 
In optical frequencies, the continuum flux decrease 0.5$\pm$0.2\% yr$^{-1}$
at 5000 \AA (Smith 2003). In this case, we obtained directly from the model
a flux decrease of 0.3\% yr$^{-1}$. The evolution of both luminosities as extracted from our model 
is in agreement with observations.

\section{Approximate  models}

Apart from the approximations we focus below,
one can also find many radiative approximations too in PWNe models: using a priori guesses for which field is dominant in each environment, using mono-chromatic assumptions for synchrotron and inverse Compton, or using Thompson cross section instead of Klein-Nishina. These assumptions certainly simplify the treatment, but at the expense of assuming approximations for which their impact is usually not checked. We have not adopted any of them here.

Regarding the diffussion-loss equation, the most usual approximation is to 
neglect the escape term (see, e.g. Tanaka \& Takahara 2010, 2011),
to obtain an advective differential equation (ADE). Using just this approximation in our complete model would lead to very similar values for the magnetic field and magnetic fraction (needed to obtain a good fit for today's Crab nebula, when imposing a correct contribution of the SSC such that it fits the high energy data today). This is because the Bohm timescale is larger than the age of the Crab nebula
and is not affecting strongly the particles' evolution.
Another common (and additional) approximation is neglecting the treatment of energy losses and instead replace it by 
the particle's escape time
(see, e.g., Zhang et al. 2008; Qiao \& Fang 2009). In this case, Eq. (\ref{te}) has the form
\begin{equation}
\label{tde}
\frac{\partial N(\gamma,t)}{\partial t}=-\frac{N(\gamma,t)}{\tau(\gamma,t)}+Q(\gamma,t),
\end{equation}
where
$\tau(\gamma,t)={\gamma}/{|\dot{\gamma}(\gamma,t)|}$ is the escape time of the particles.
In this case, particles are not losing energy, but they are rather removed from the distribution after a certain time. This makes Eq. (\ref{tde}) a partial differential equation in
time only (TDE).

Before doing the fits, we fixed the parameters which are obtained from observations, as we have done in the complete model,
and included the ADE and TDE cases with the complementary approximations done by Tanaka \& Takahara (2010, hereafter ADE-T) and Zhang et al. (2008,
hereafter TDE-Z). 
In ADE-T, the bremsstrahlung energy losses and its spectrum, and the FIR and NIR contributions into the inverse Compton energy losses and their spectrum, 
are ignored. Also, the maximum energy at injection is fixed and the expansion of the PWN is modeled in a ballistic approximation ($R_{PWN}=v_{PWN}t$).
All these approximations are not done in the full treatment presented above, against which we compare.
In the TDE-Z, only the synchrotron escape time is considered (thus ignoring all other timescales) and Bohm diffusion is used.

Table~\ref{apT} shows the parameters for  each of the 
models needed  to obtain a  good fit of the Crab Nebula data at the current age.  
The column labelled {\it value} 
are the parameters of the complete model of \S 2 (the origin of each parameter is commented in Table \ref{crabconstrains}). The dots appear when no change is needed from those values.

Given that the observational parameters such as the age, the breaking index, 
the period, and the period derivative are fixed, they continue 
to determine $\tau_0$ and $\tau_c$ in all models. 
For the TDE model, the break energy and the shock radius fraction (and, in consequence, the maximum energy at injection) have decreased. A smaller shock radius diminishes the number of VHE electrons, which is necessary due to the lack of energy losses affecting the population, and the smaller energy break corrects the lack of radio flux. 
The initial spin-down luminosity is smaller because the lack of losses makes electrons' disappearance slower. The magnetic field fraction is larger to power
the SSC contribution, and the energy break increases to compensate the lack of escaping particles at low energies and correct the radio flux.
In the ADE-T case, we take the expansion velocity given by Tanaka \& Takahara (2010) of 1800 km s$^{-1}$, which gives a radius for the PWN of 1.7 pc. This means that the synchrotron radiation is confined in a smaller volume, so the synchrotron photon density is  larger
and the magnetic energy fraction needed to obtain the correct SSC contribution is smaller than in our case. Note that the minimum and maximum energy at injection are also fixed in
time, according with the values used in Tanaka and Takahara (2010).

\begin{table}
\centering
\scriptsize
\caption{Comparison of the values used or obtained in the different fits
of the  Crab Nebula today. Meaning and units of variables are as in Table 1. We use dots for
those parameters which have the same values as in the complete model.}
\vspace{0.2cm}
\label{apT}
\begin{tabular}{l lll}
\hline
 Symbol & Value &   {ADE-T} & {TDE-Z}\\
\hline
  $L(t_{age})$ & $4.5 \times 10^{38}$  & $\ldots$ & $2.5 \times 10^{38}$\\
\hline
  $\gamma_{min}(t)$ & 1 & $10^2$ & \ldots\\
  $\gamma_{max}(t)$ & $7.9 \times 10^9$  & $7 \times 10^9$ (fixed) & $6.5 \times 10^9$\\
  $\gamma_b$ & $7 \times 10^5$  & $7 \times 10^5$ & $9 \times 10^5$\\
  $\alpha_1$ & 1.5 & \dots & \dots\\
  $\alpha_2$ & 2.5 & $\ldots$ & \ldots\\
  $\varepsilon$ & $1/3$  & \ldots & \ldots\\
\hline
 $L_0$ & $3.1 \times 10^{39}$ & $\ldots$ & $1.7 \times 10^{39}$ \\
 $B(t_{age})$ & 97  & $\ldots$ & 93\\
 $\eta$ & 0.012 &  0.006 & 0.015\\
 $R_{PWN}$ & 2.1 & 1.7 & 1.9\\
\hline
 $T_{CMB}$ & 2.73  & $\ldots$ & $\ldots$\\
 $w_{FIR}$ & 0.25  & $\ldots$ & $\ldots$\\
 $T_{FIR}$ & 70 & 0 & \ldots\\
 $w_{FIR}$ & 0.5  & 0 & \ldots\\
 $T_{NIR}$ & 5000 & 0 & \ldots\\
 $w_{NIR}$ & 1  & 0 & \dots\\
 $n_H$ & 1 & 0 & \ldots\\
\hline
\hline
\end{tabular}
\end{table}

\begin{figure*}
\begin{center}
\begin{tabular}{lr}
\includegraphics[scale=0.45]{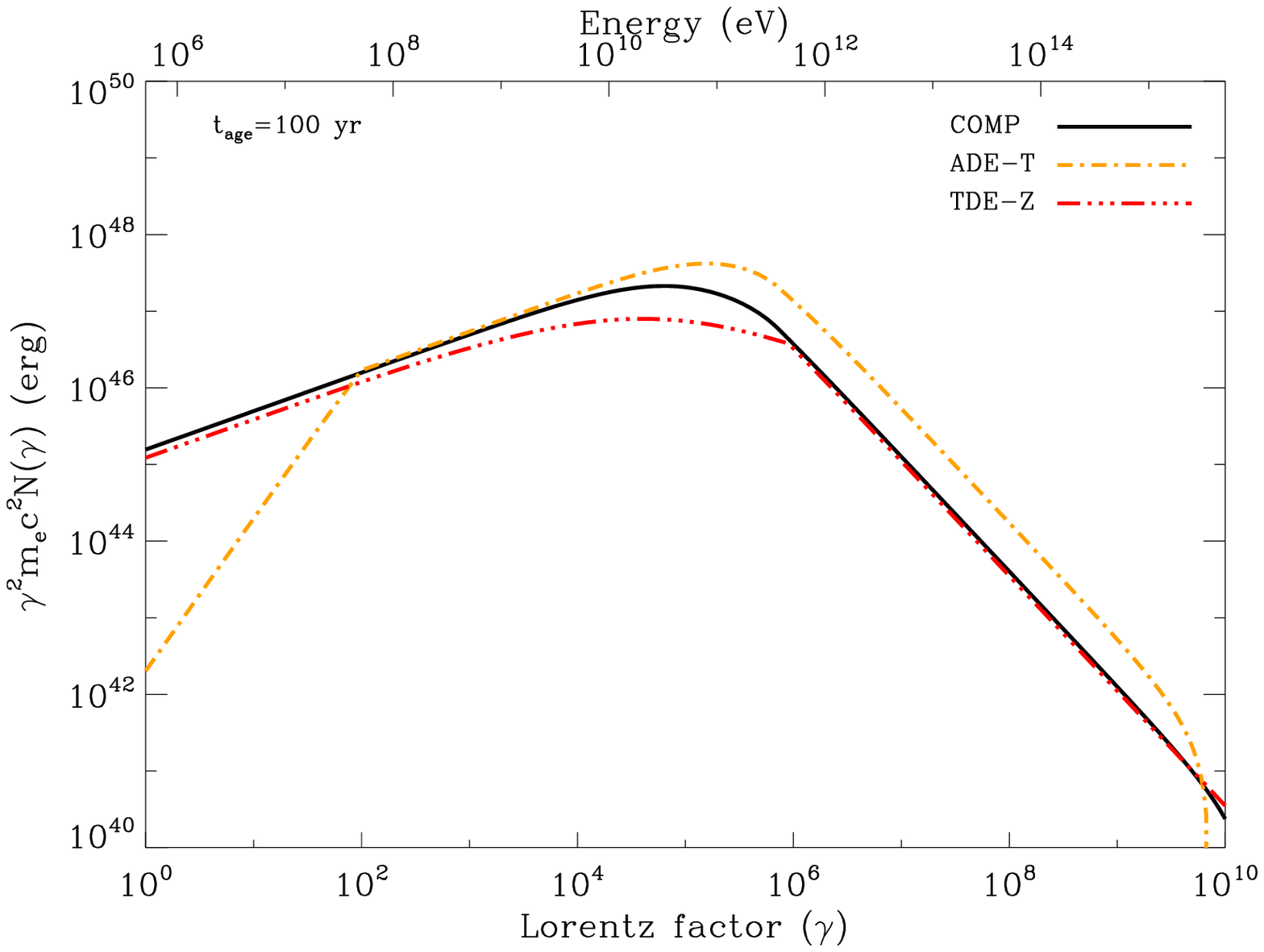}  
\includegraphics[scale=0.45]{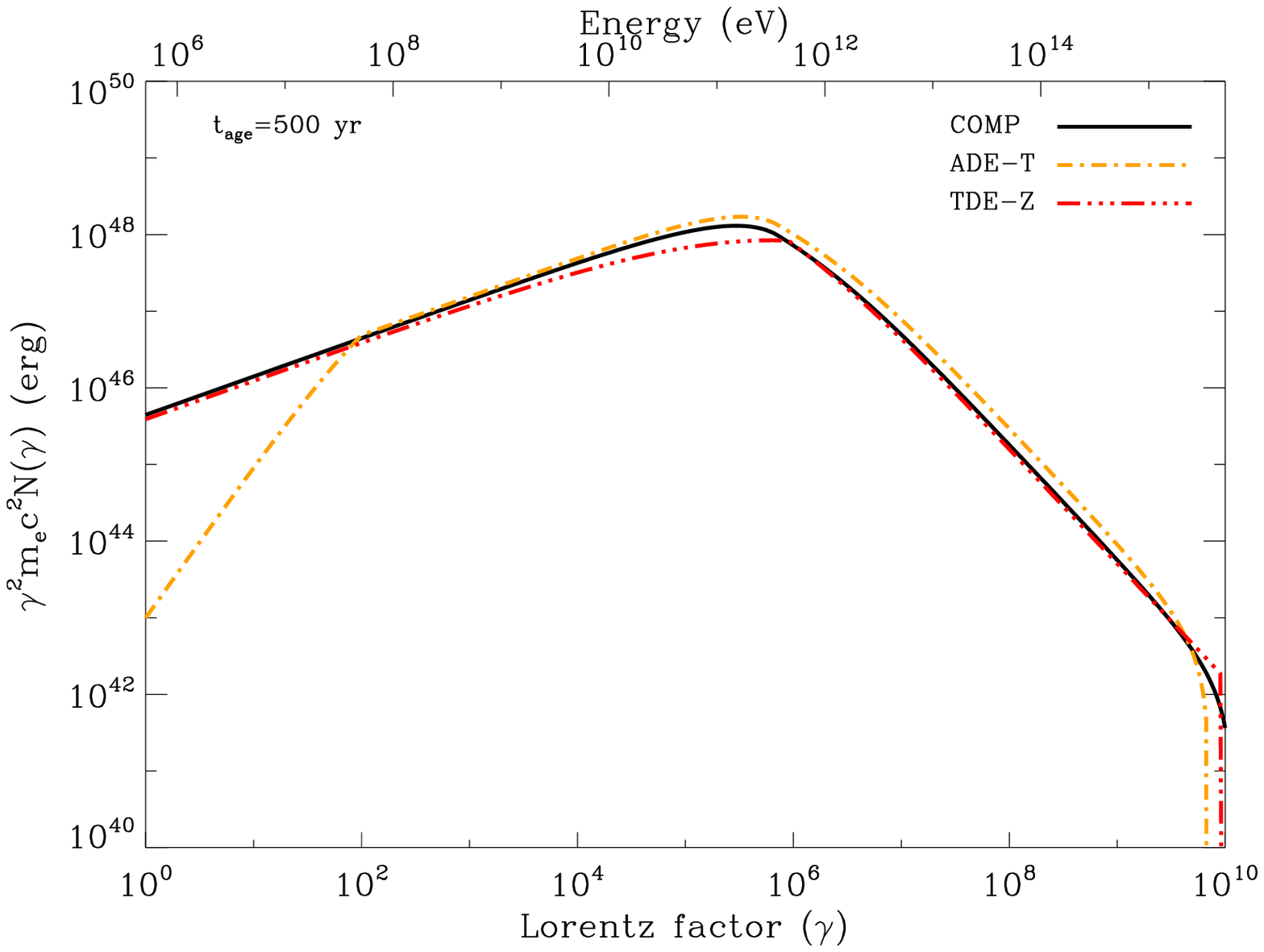}\\
\includegraphics[scale=0.45]{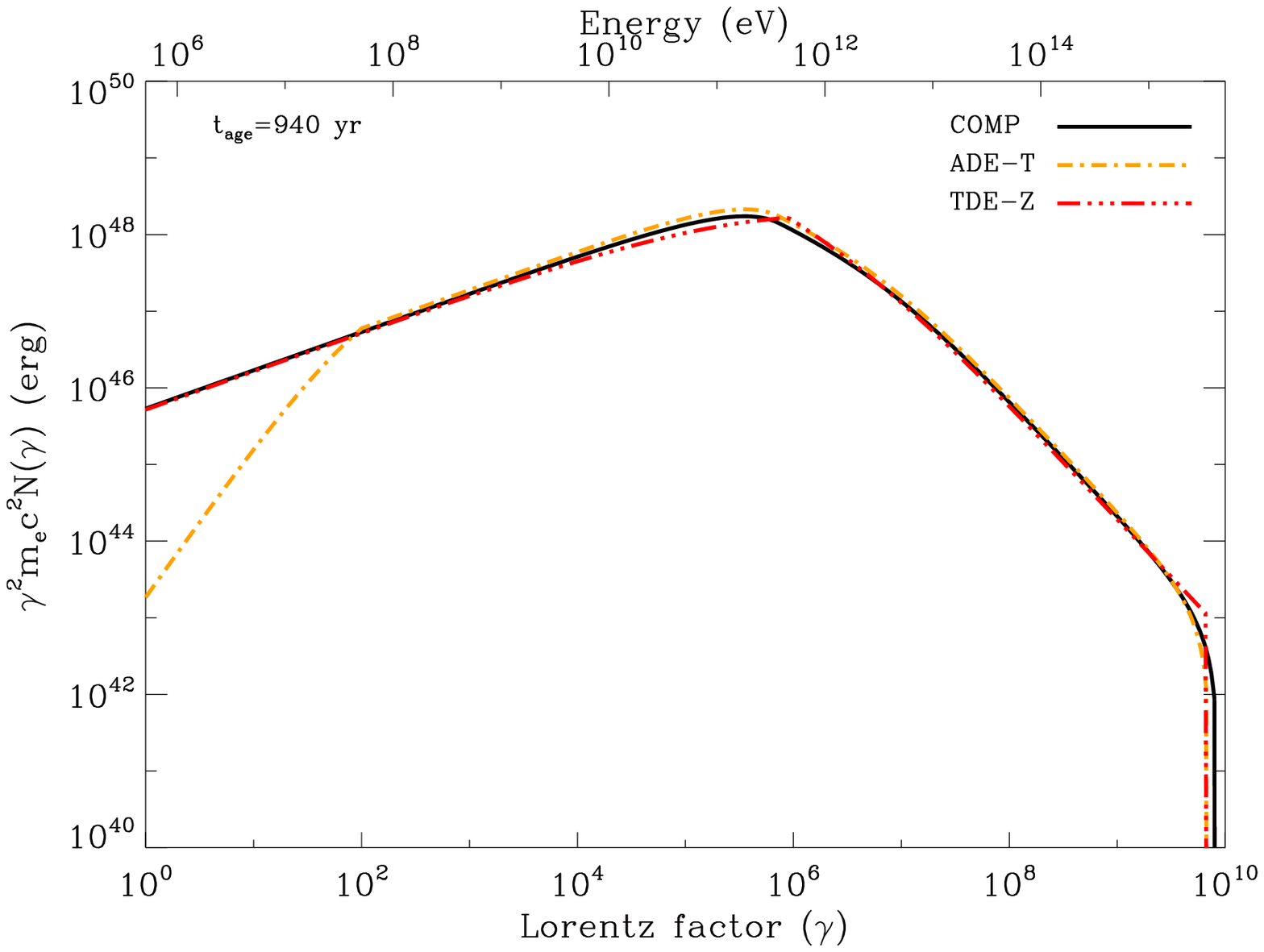} 
\includegraphics[scale=0.45]{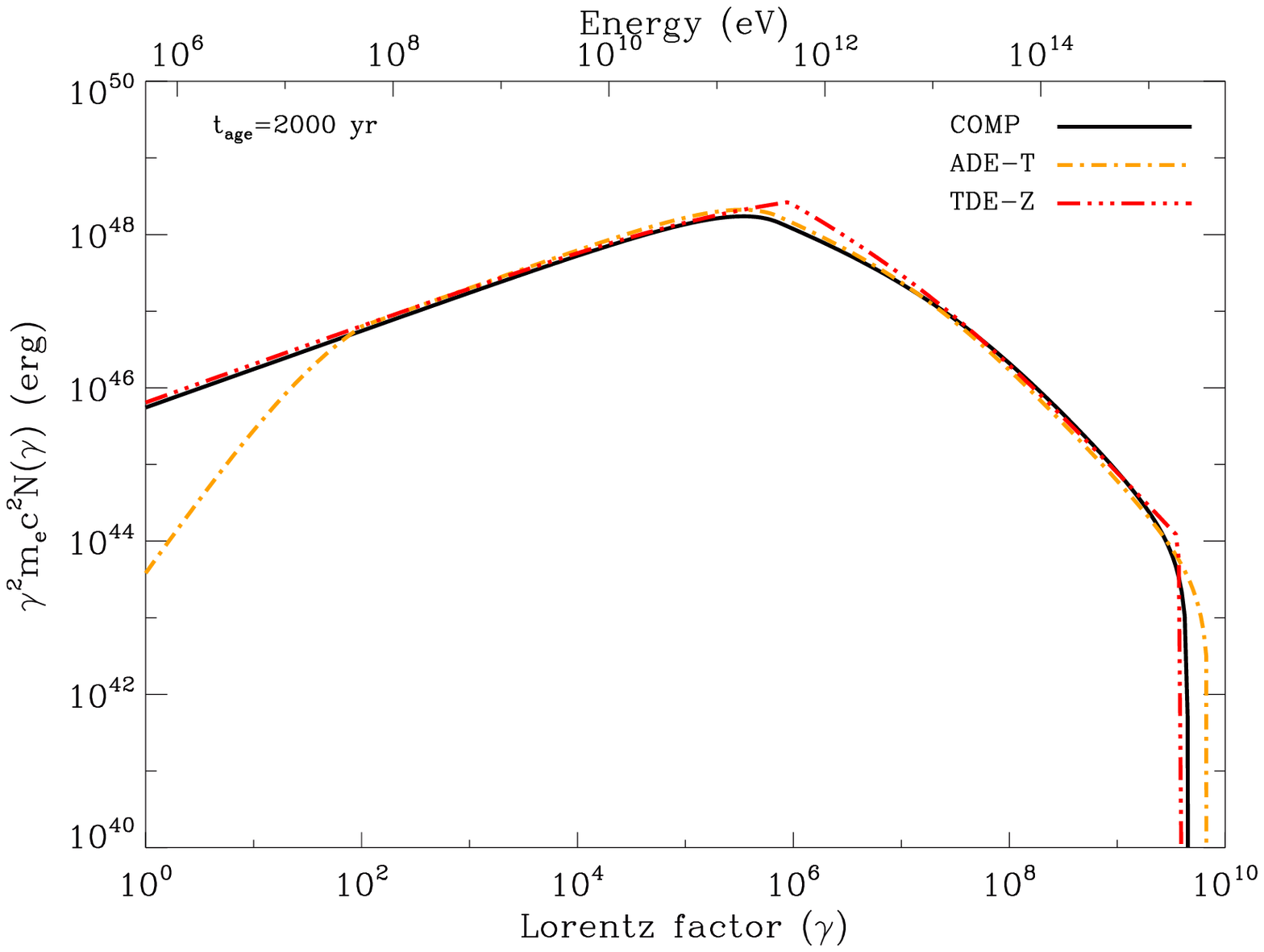}\\
\includegraphics[scale=0.45]{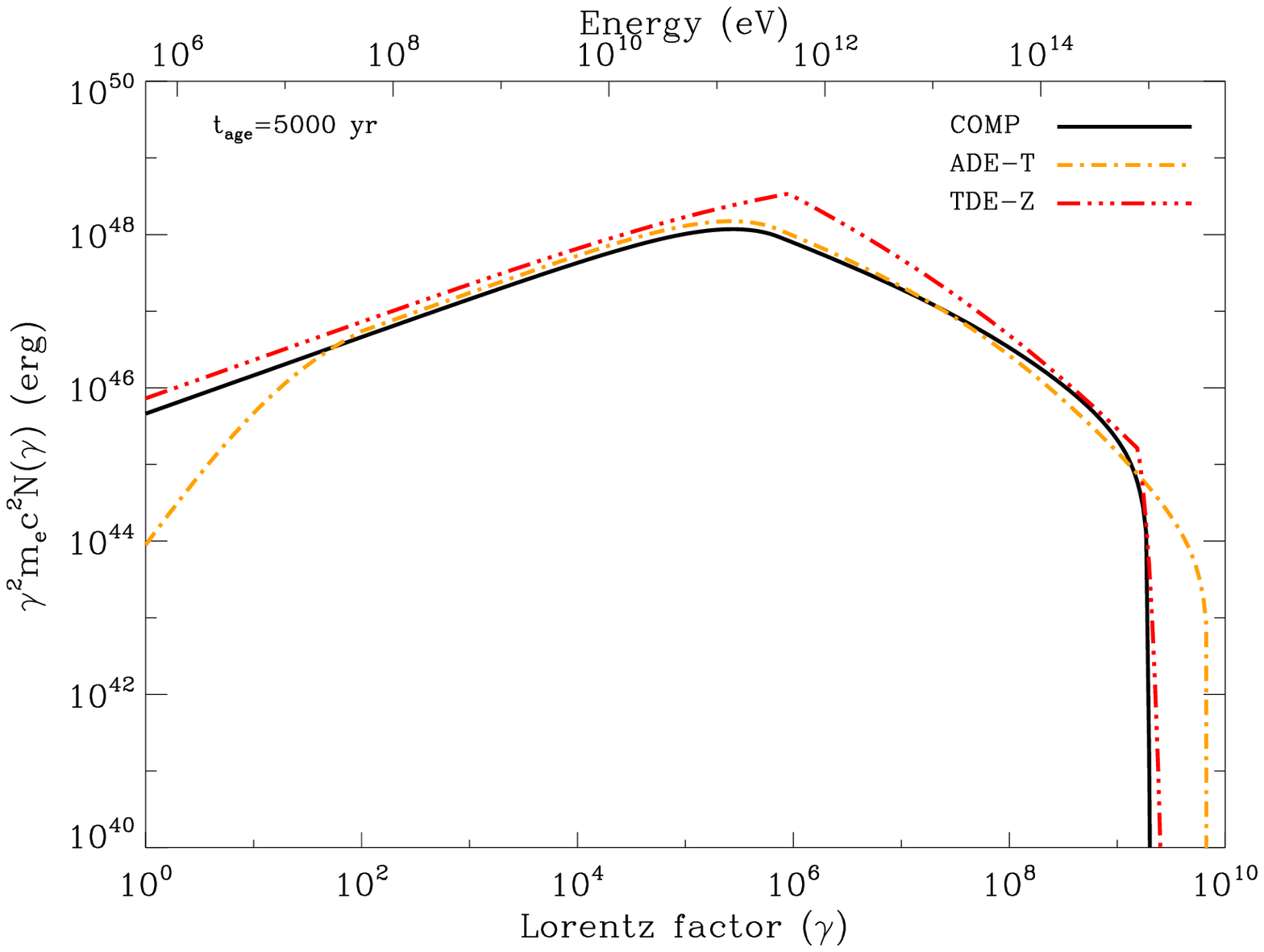} 
\includegraphics[scale=0.45]{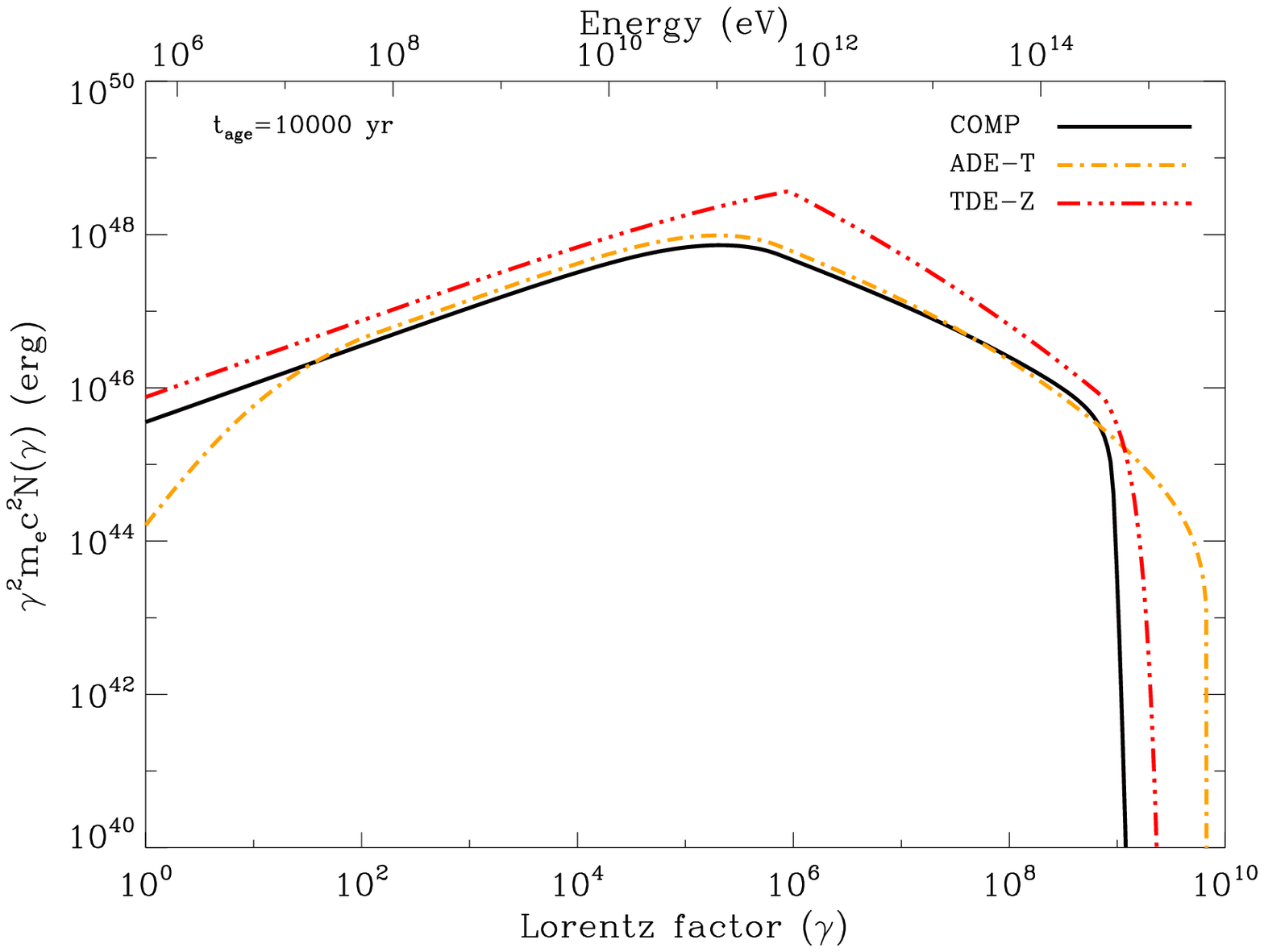}
\end{tabular}
\end{center}
\caption{Electron distribution of the Crab nebula 
computed for different ages using the complete model, together with the obtained results under the  ADE-T, and TDE-Z approximations.}
\label{comp-resu-elec}
\end{figure*}

\begin{figure*}
\begin{center}
\begin{tabular}{lr}
\includegraphics[scale=0.45]{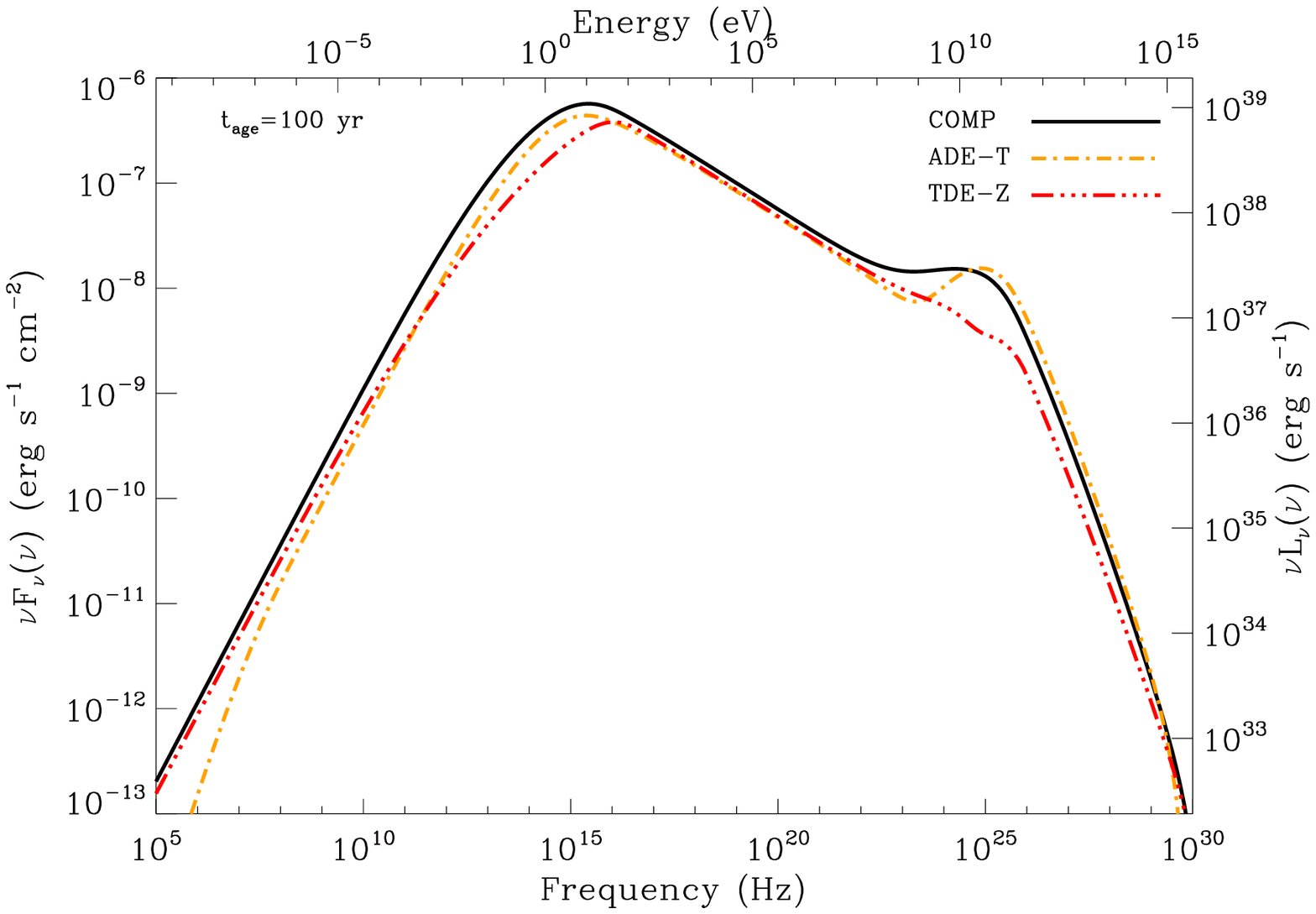}  
\includegraphics[scale=0.45]{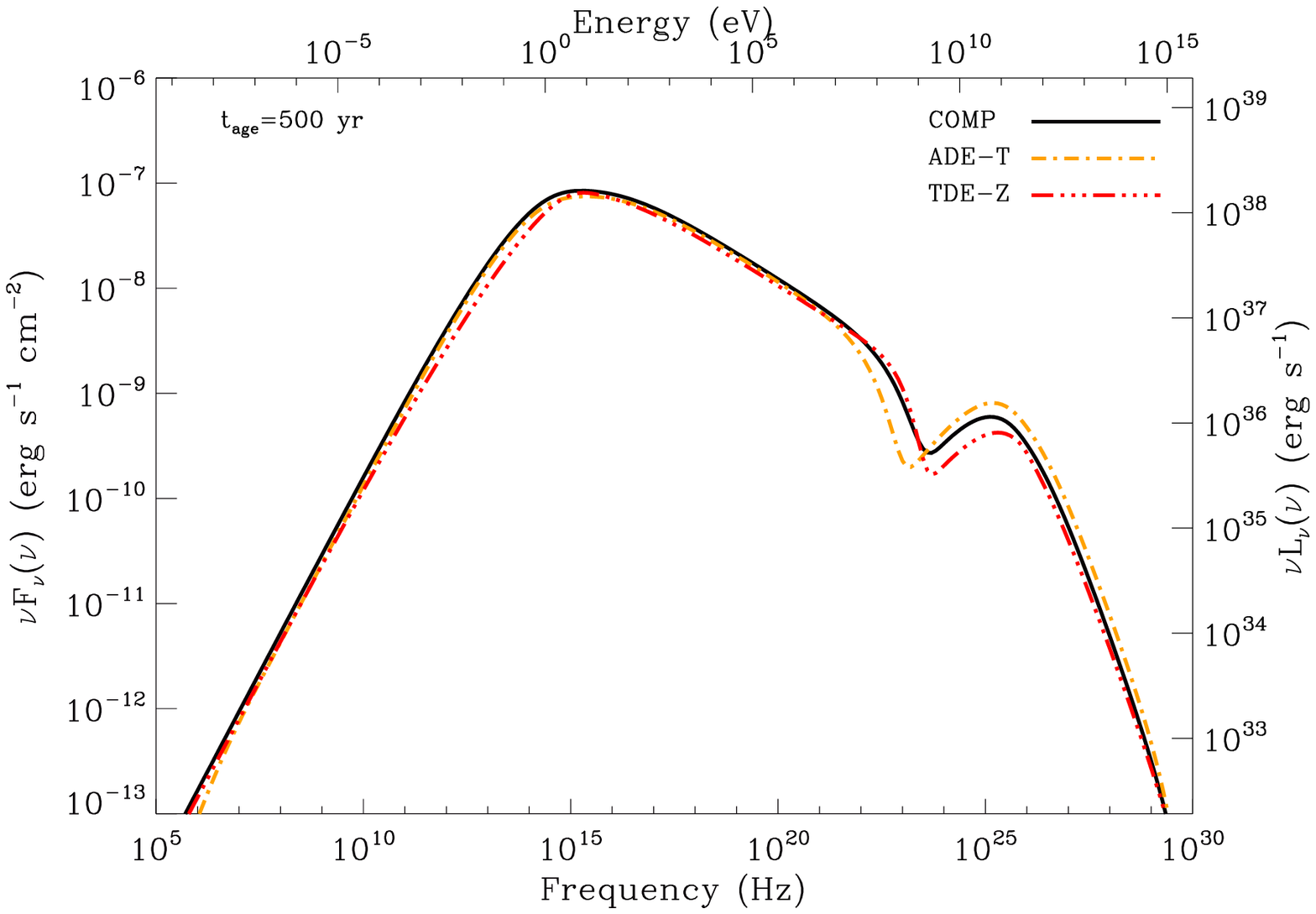}\\
\includegraphics[scale=0.45]{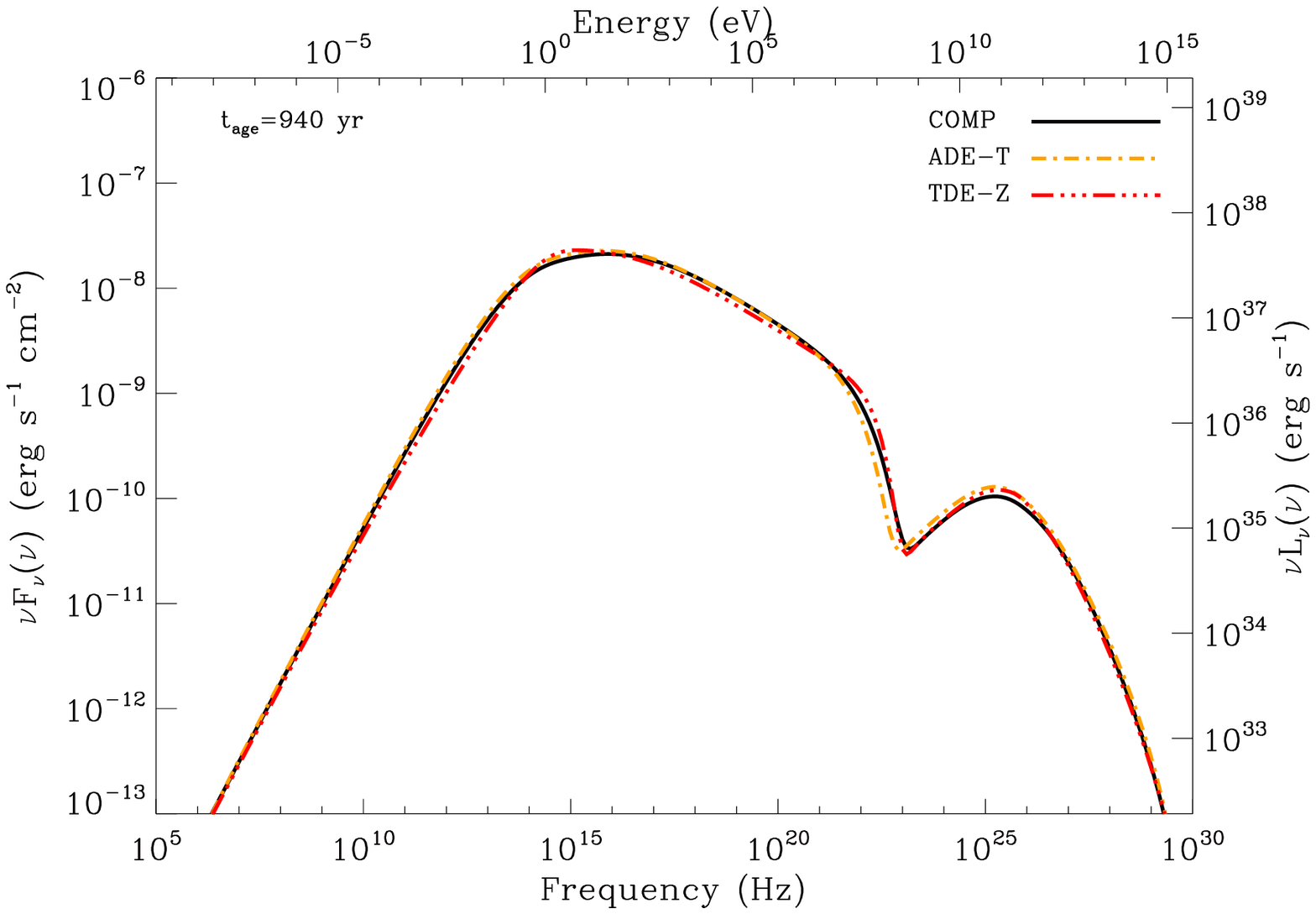}  
\includegraphics[scale=0.45]{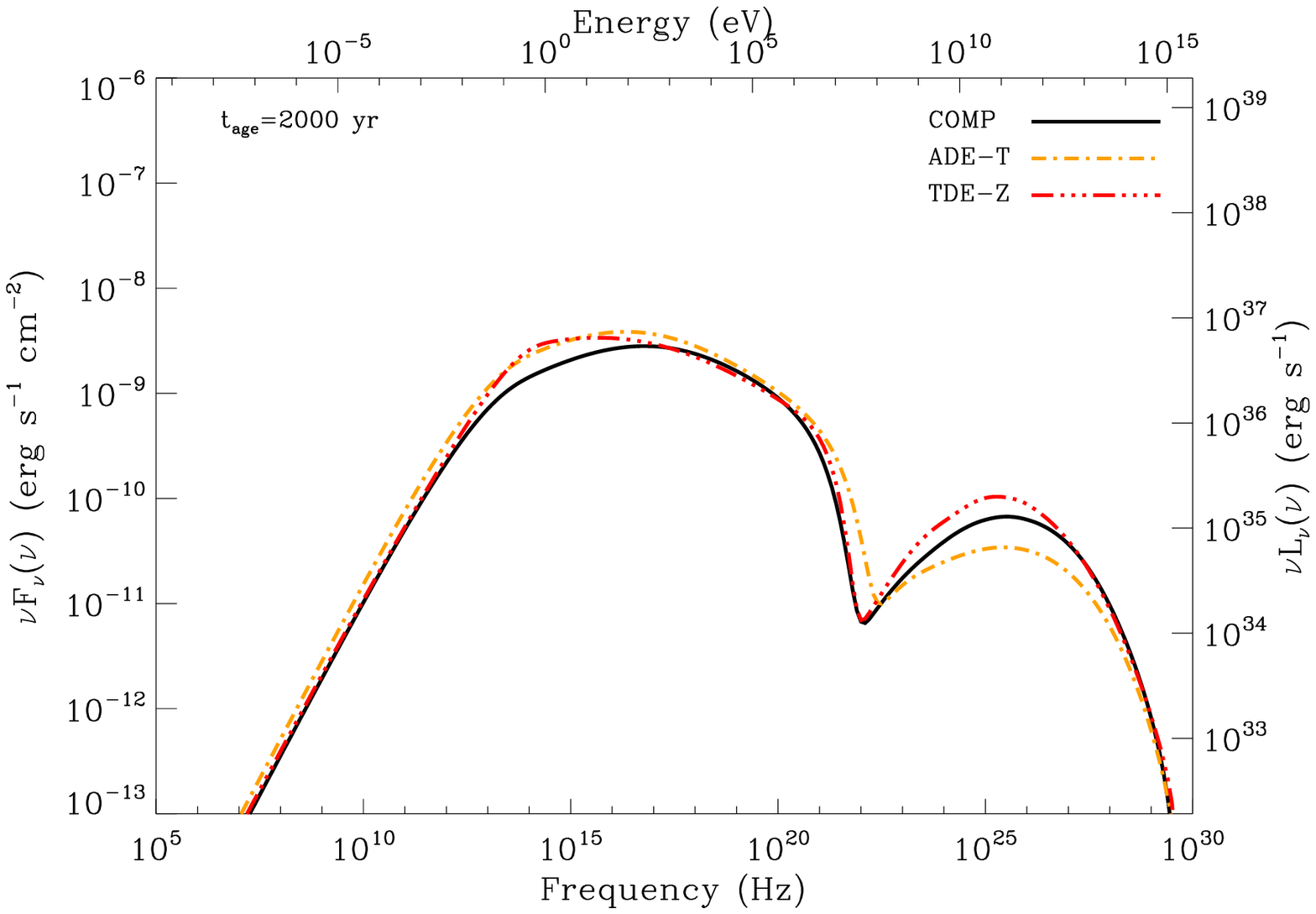}\\
\includegraphics[scale=0.45]{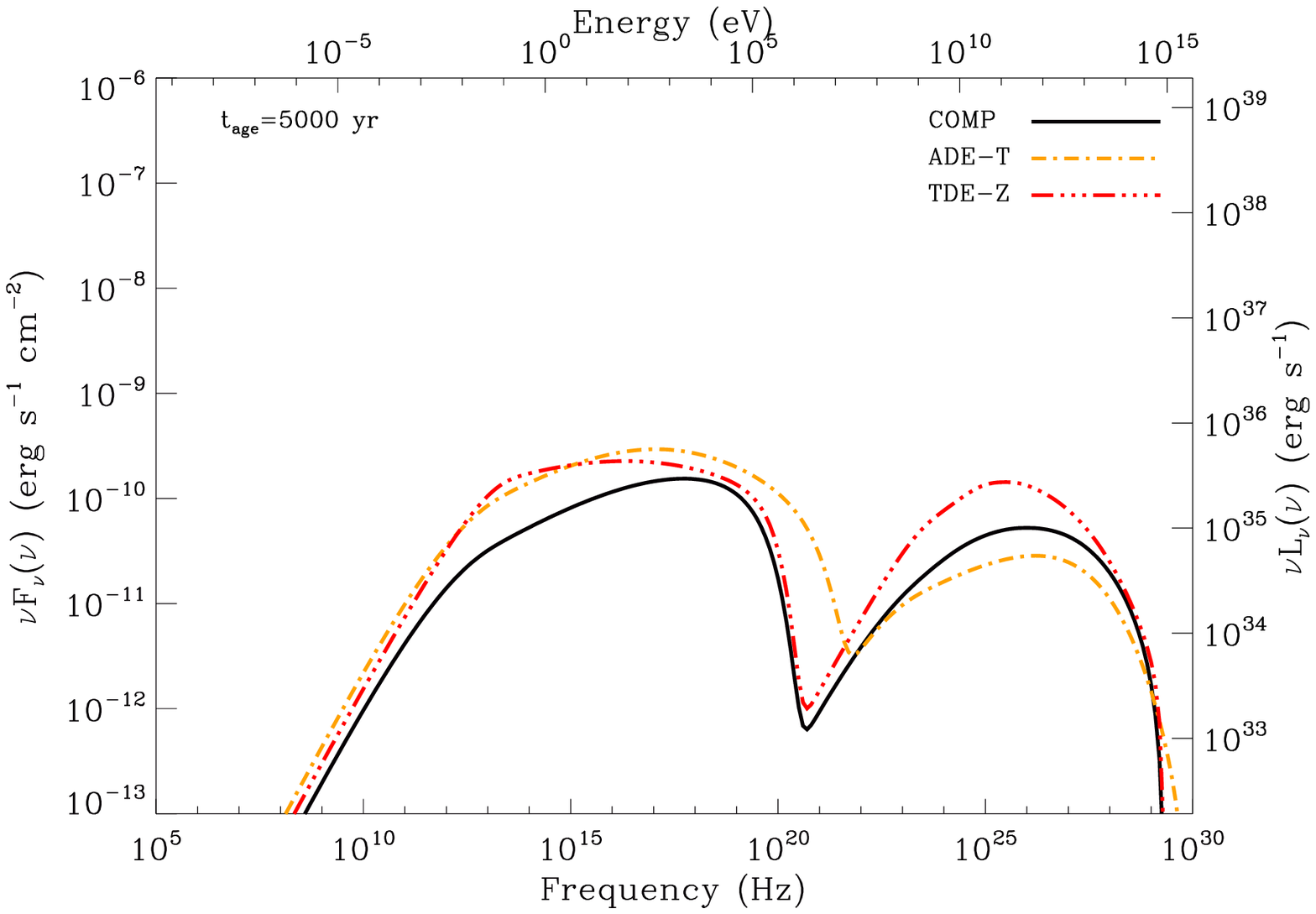} 
\includegraphics[scale=0.45]{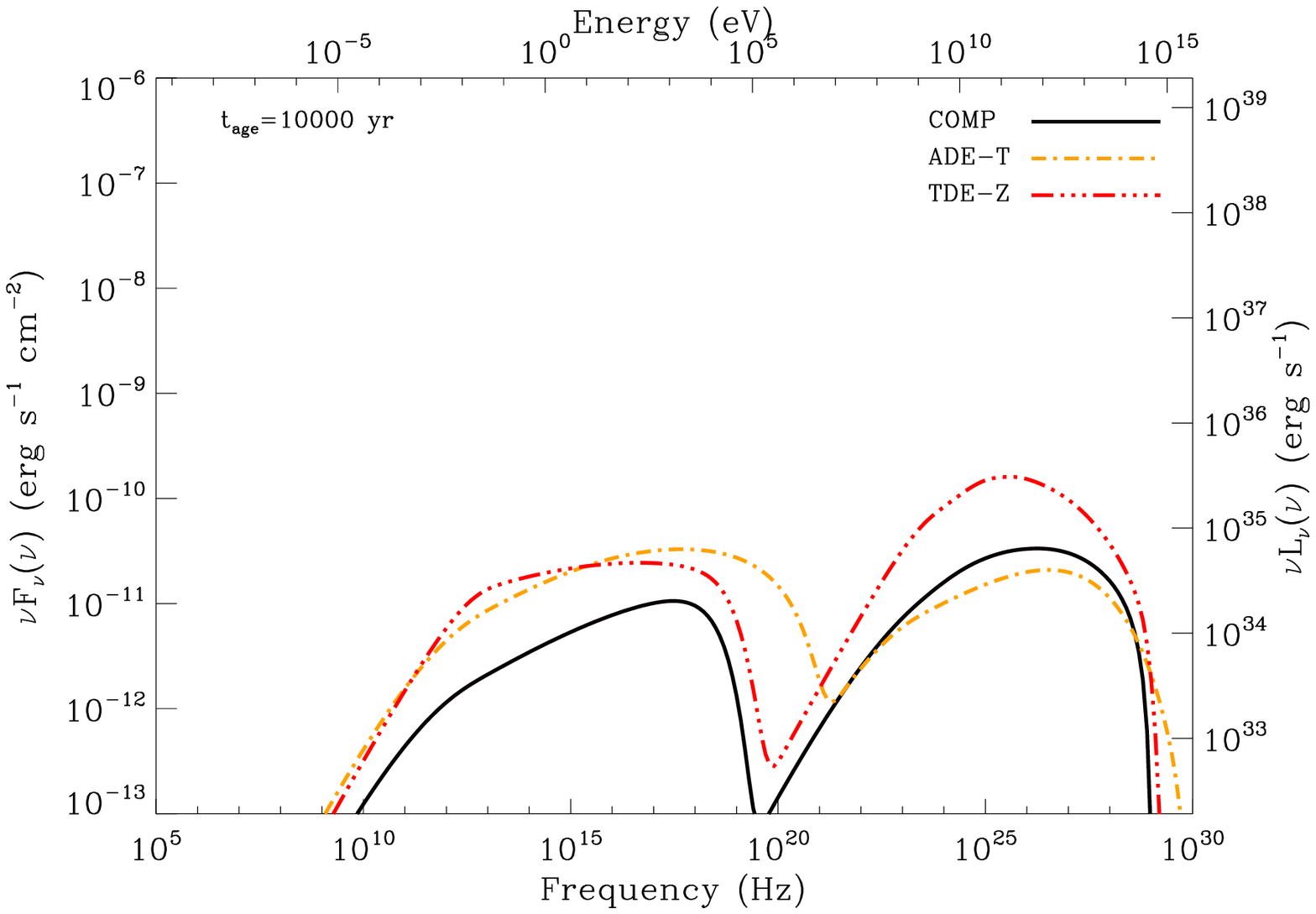}
\end{tabular}
\end{center}
\caption{Photon spectrum of the Crab nebula computed for different ages using the complete model, together with the obtained results under the  ADE-T, and TDE-Z approximations.}
\label{comp-resu-spec}
\end{figure*}

\begin{figure*}
\begin{center}
\begin{tabular}{lr}
\includegraphics[scale=0.45]{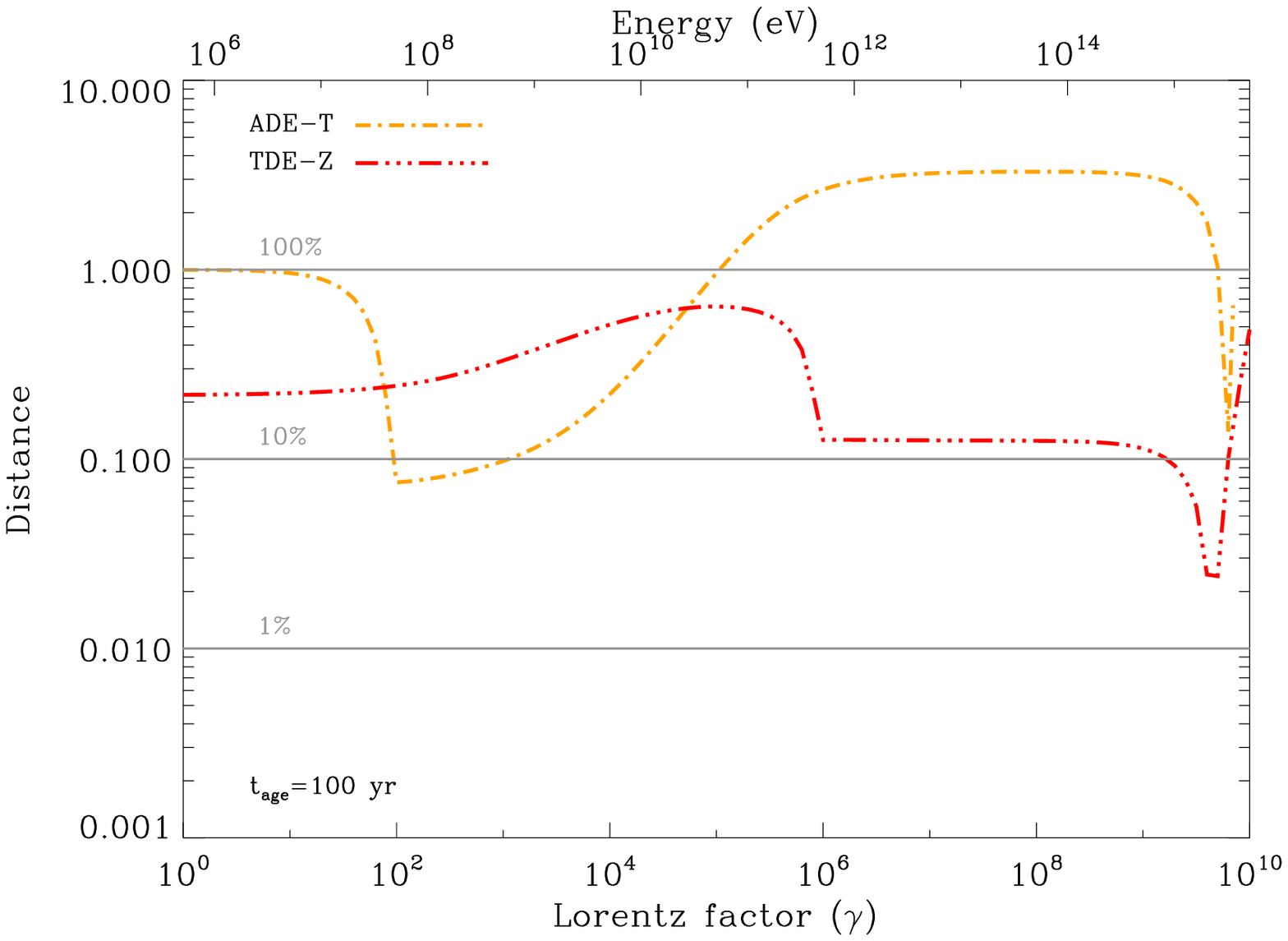}  
\includegraphics[scale=0.45]{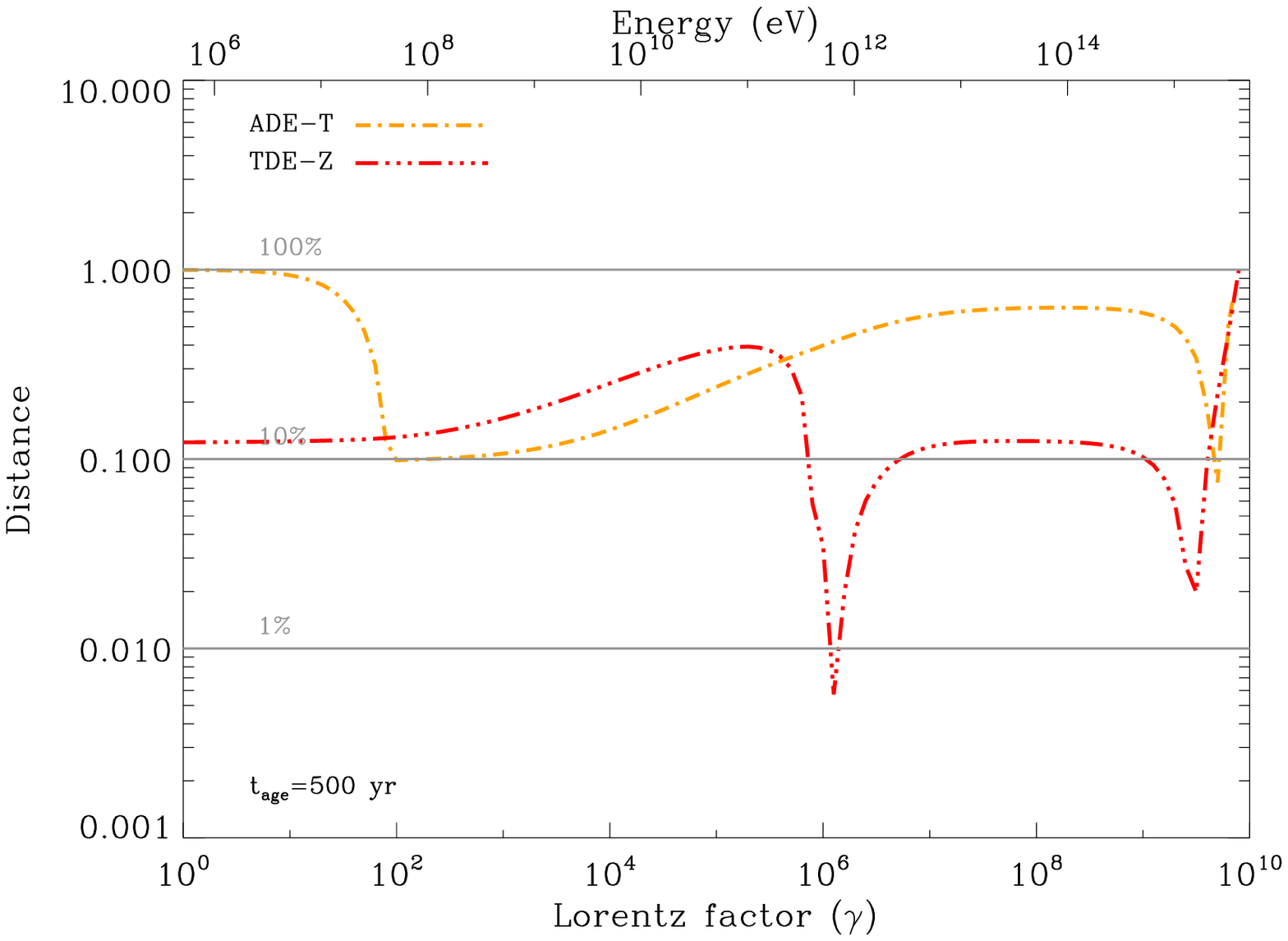}\\
\includegraphics[scale=0.45]{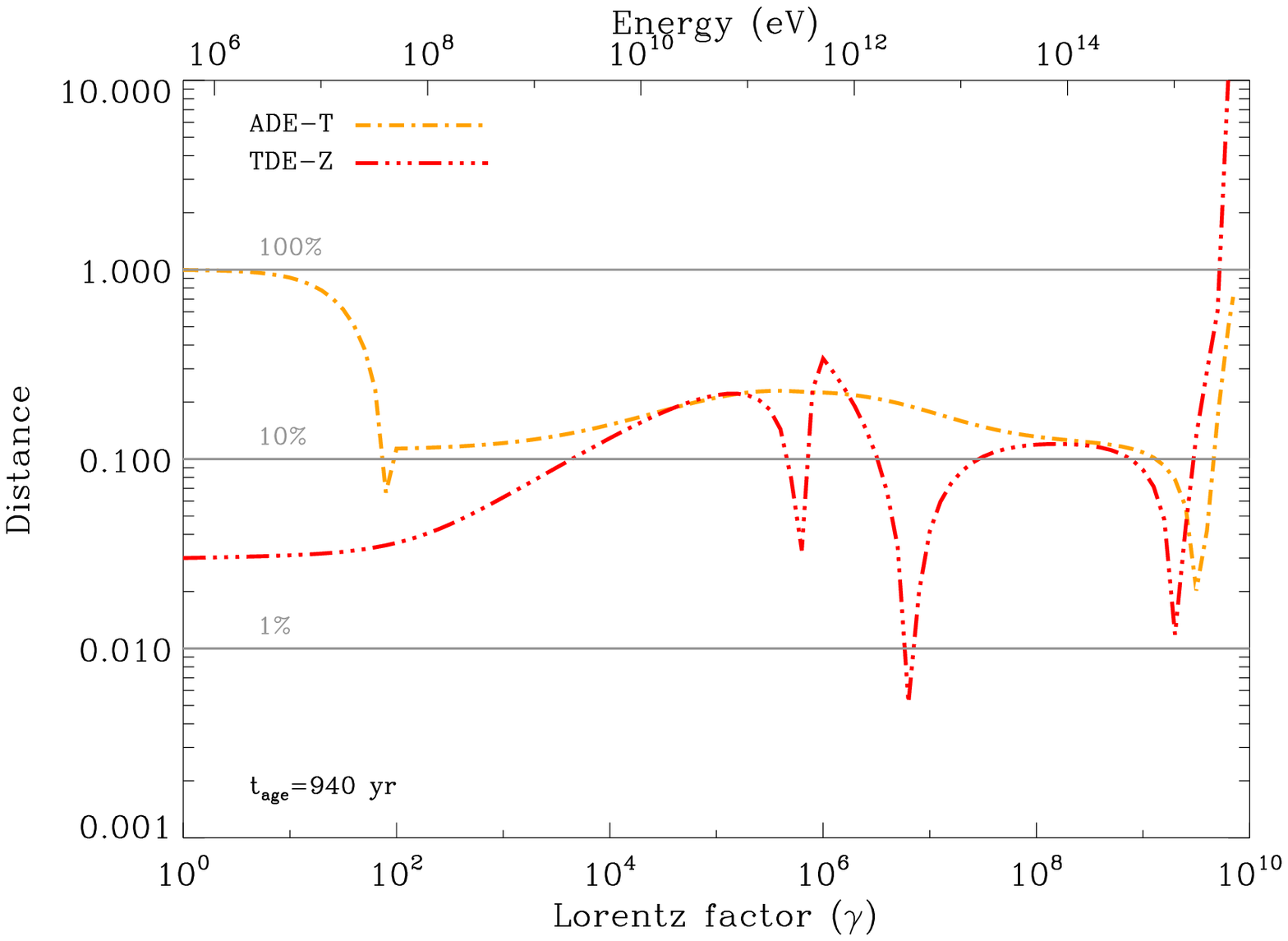} 
 \includegraphics[scale=0.45]{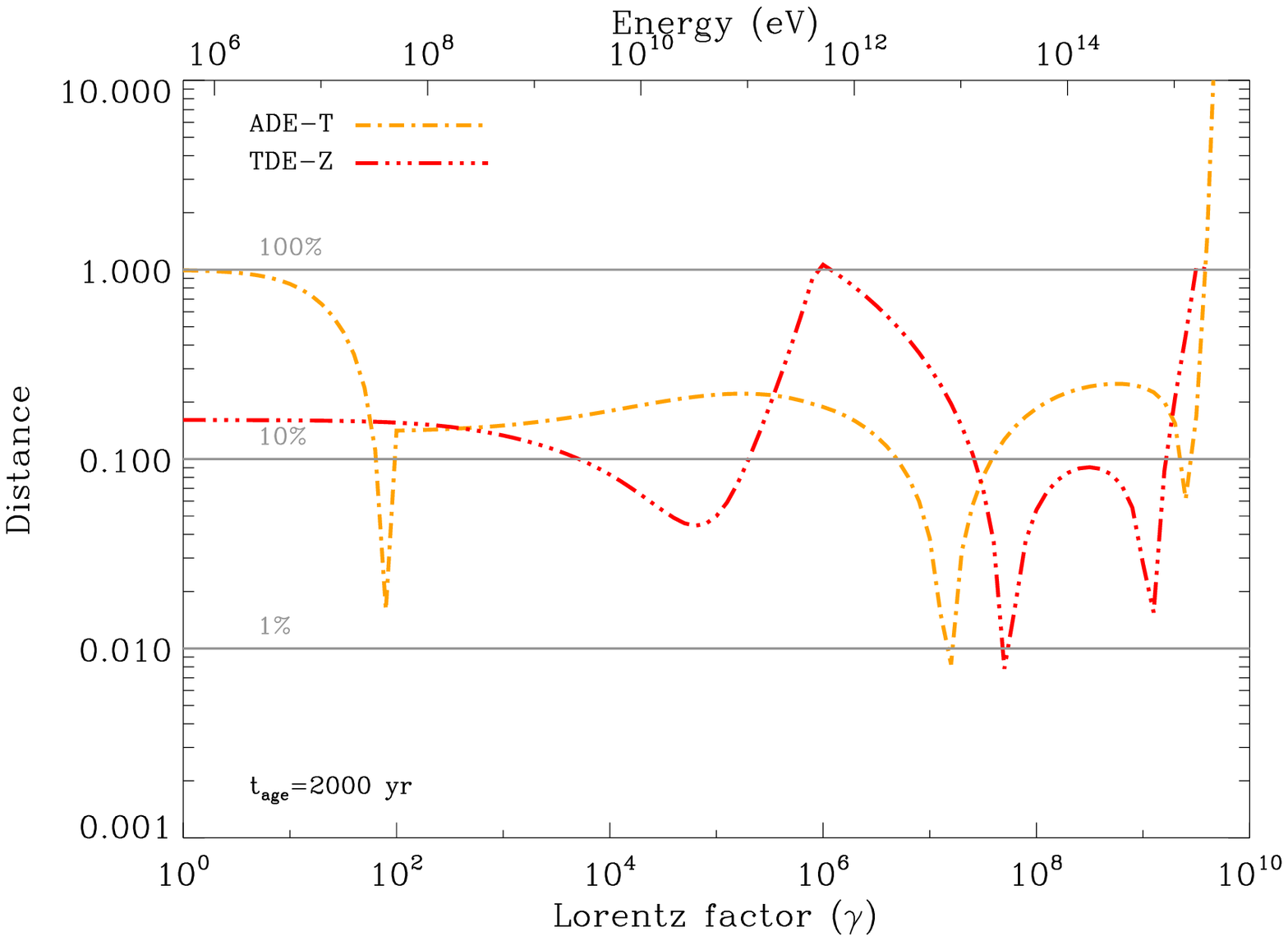}\\
\includegraphics[scale=0.45]{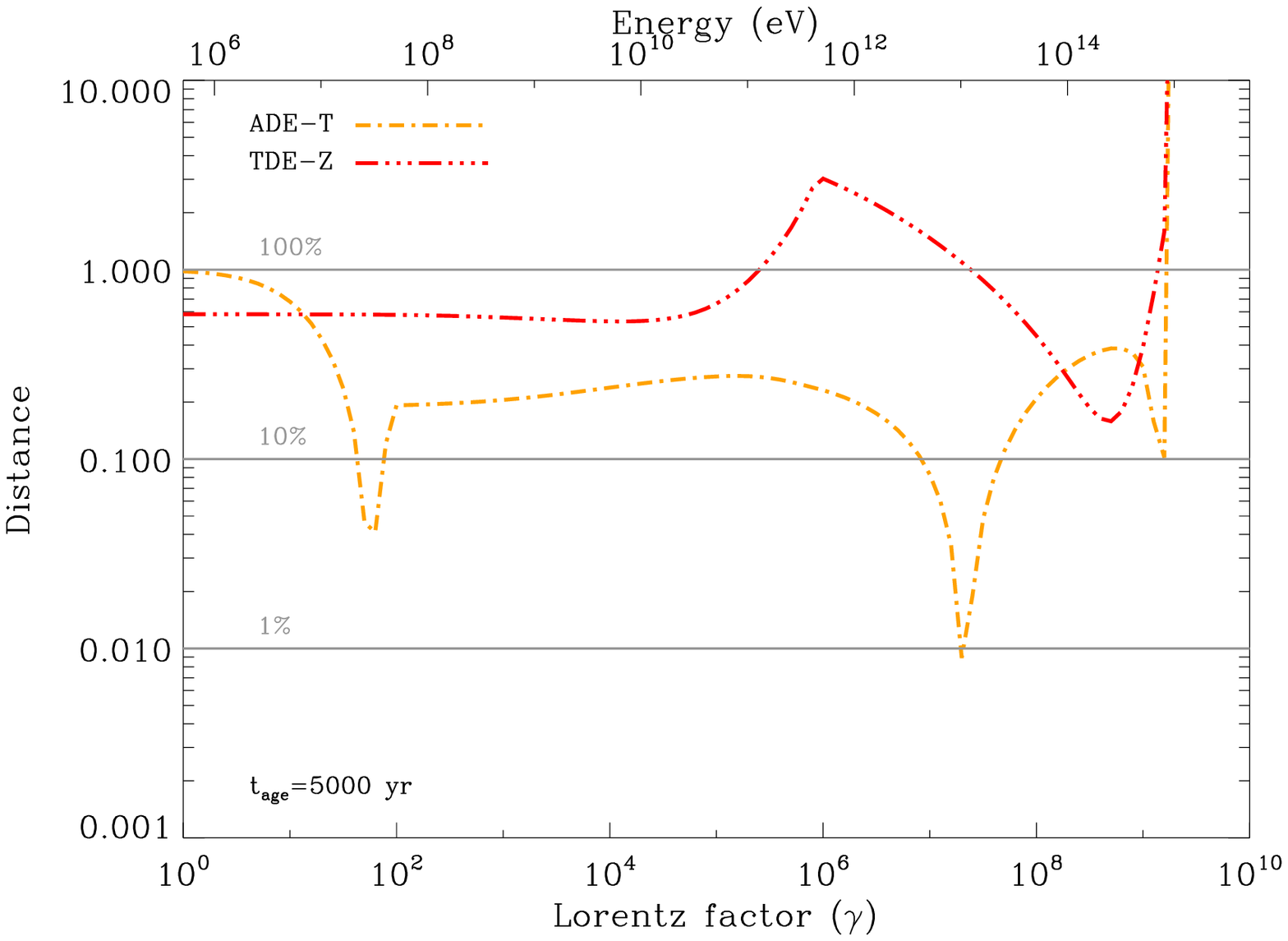} 
 \includegraphics[scale=0.45]{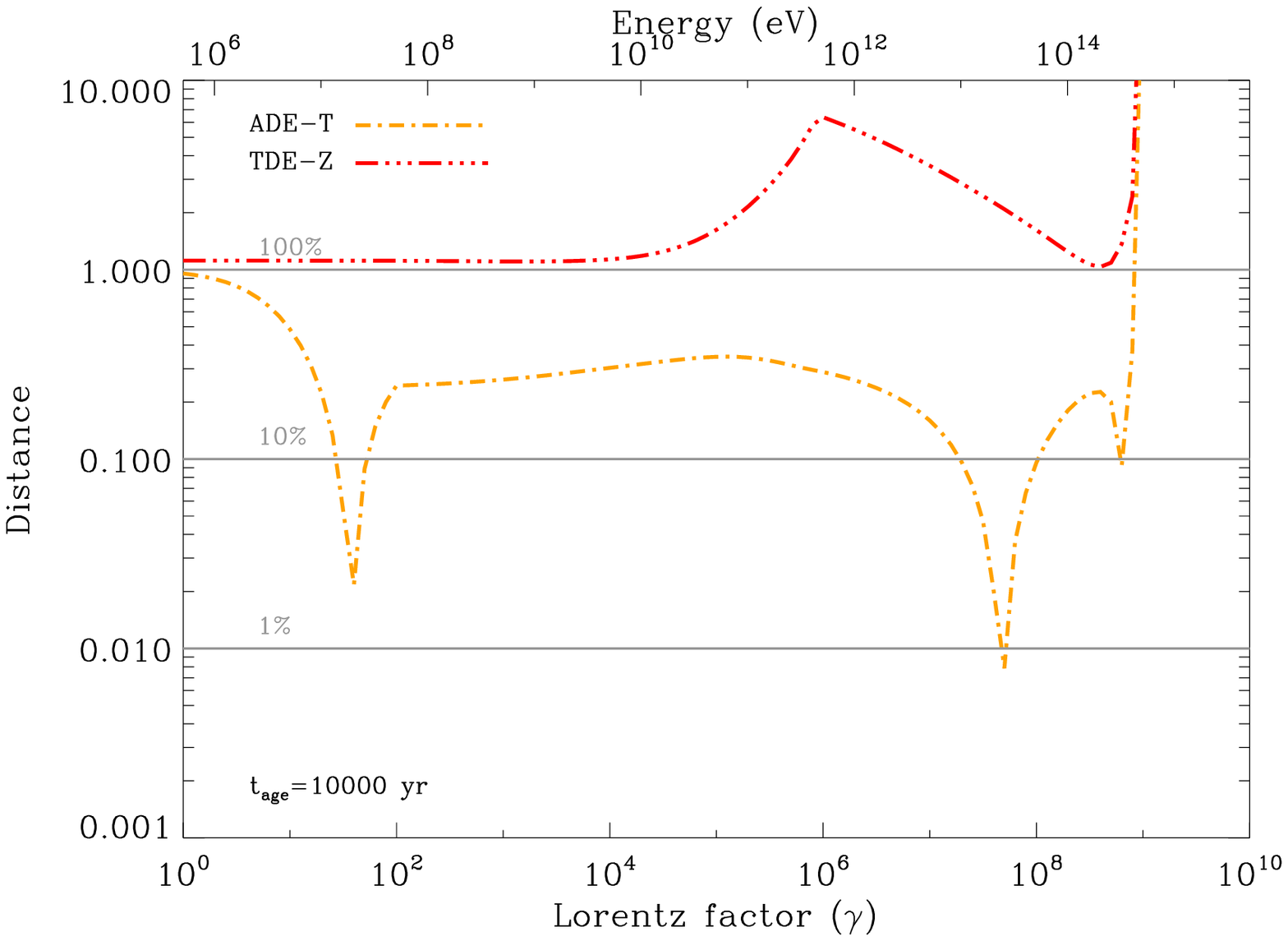}
\end{tabular}
\end{center}
\caption{Relative distance of the results for the electron distribution between the complete model, the ADE-T, and TDE-Z approximations for the Crab Nebula at different ages. }
\label{rel-elec}
\end{figure*}

\begin{figure*}
\begin{center}
\begin{tabular}{lr}
\includegraphics[scale=0.45]{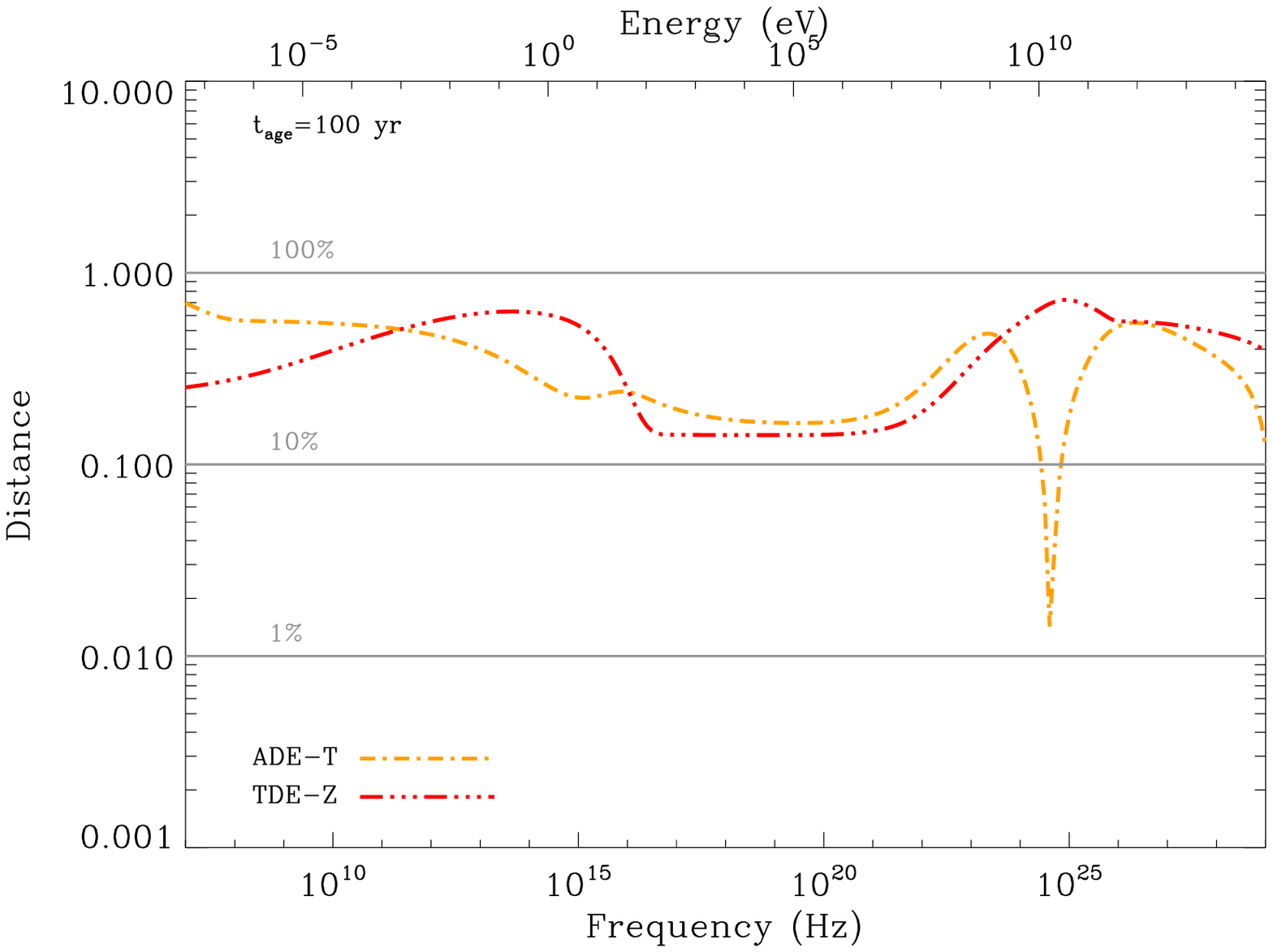}  
\includegraphics[scale=0.45]{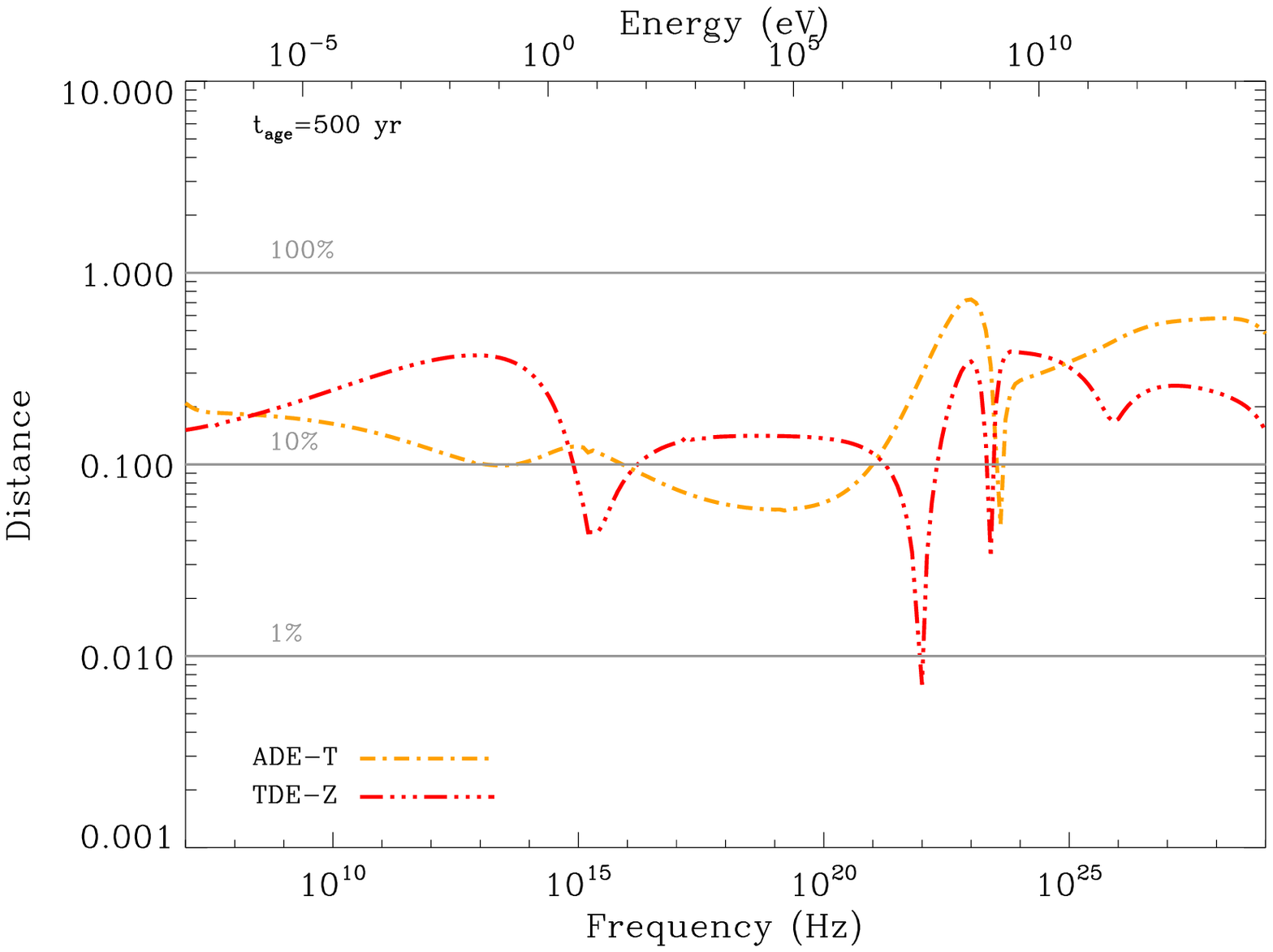}\\
\includegraphics[scale=0.45]{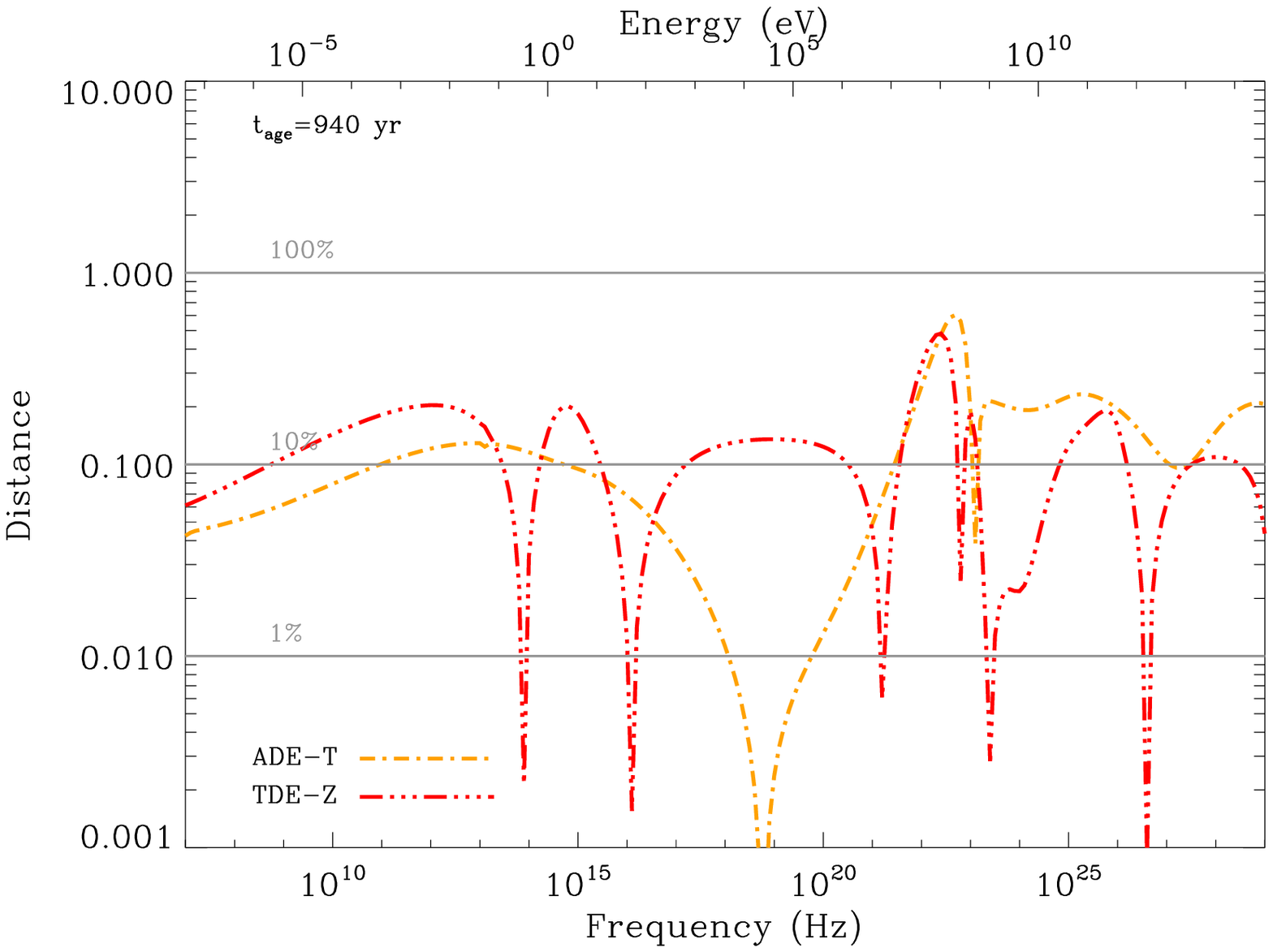}  
\includegraphics[scale=0.45]{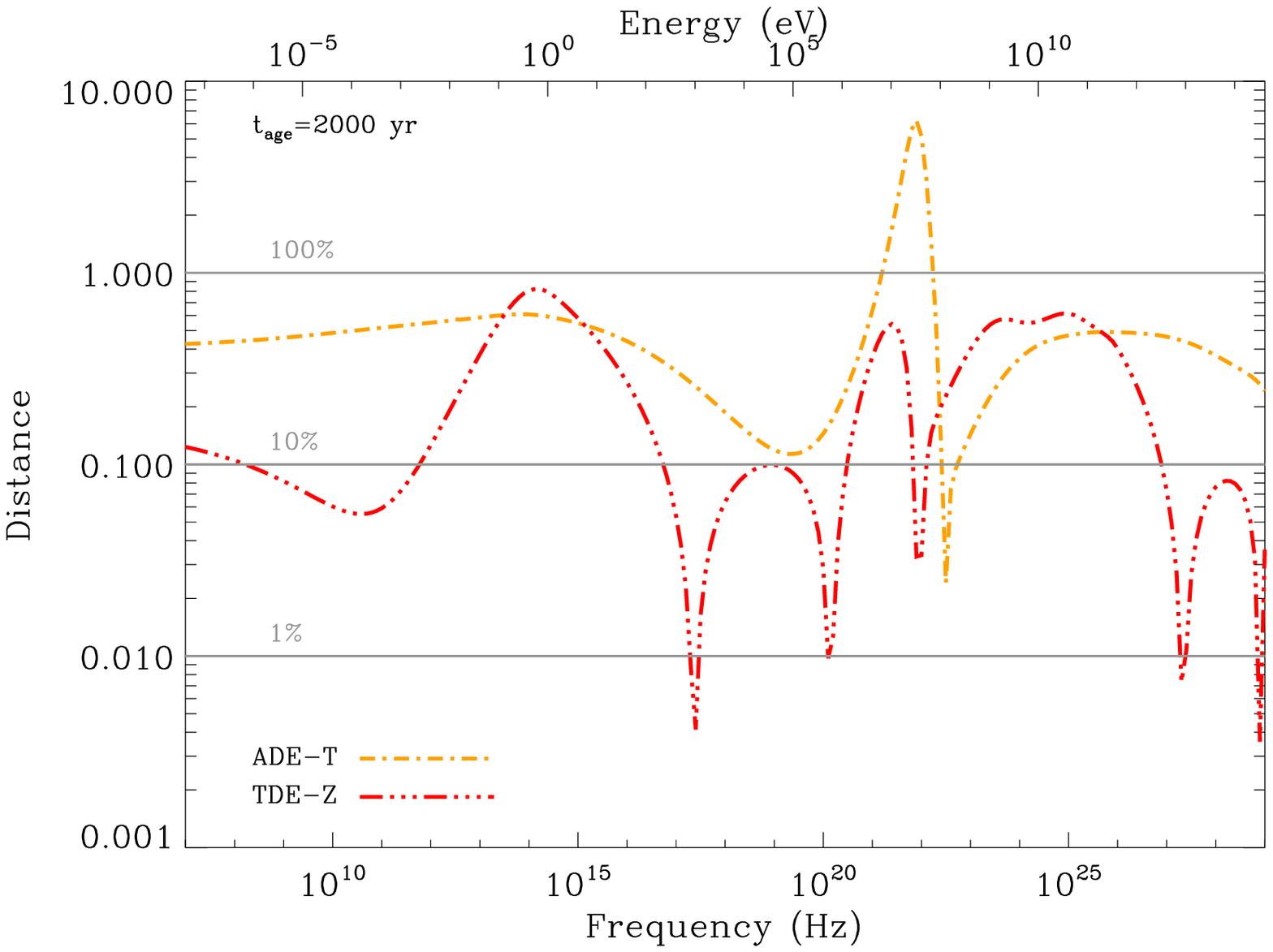}\\
\includegraphics[scale=0.45]{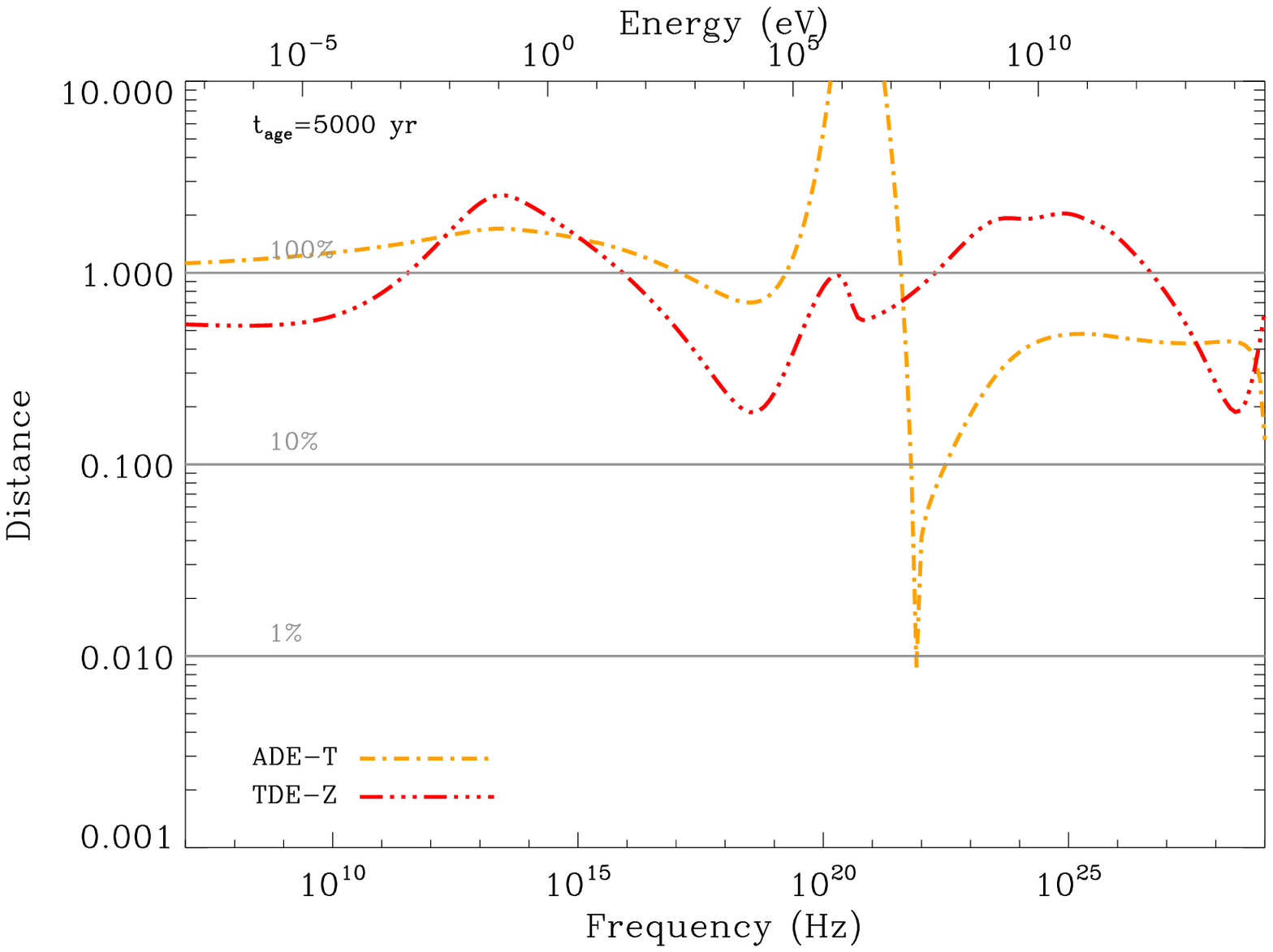} 
\includegraphics[scale=0.45]{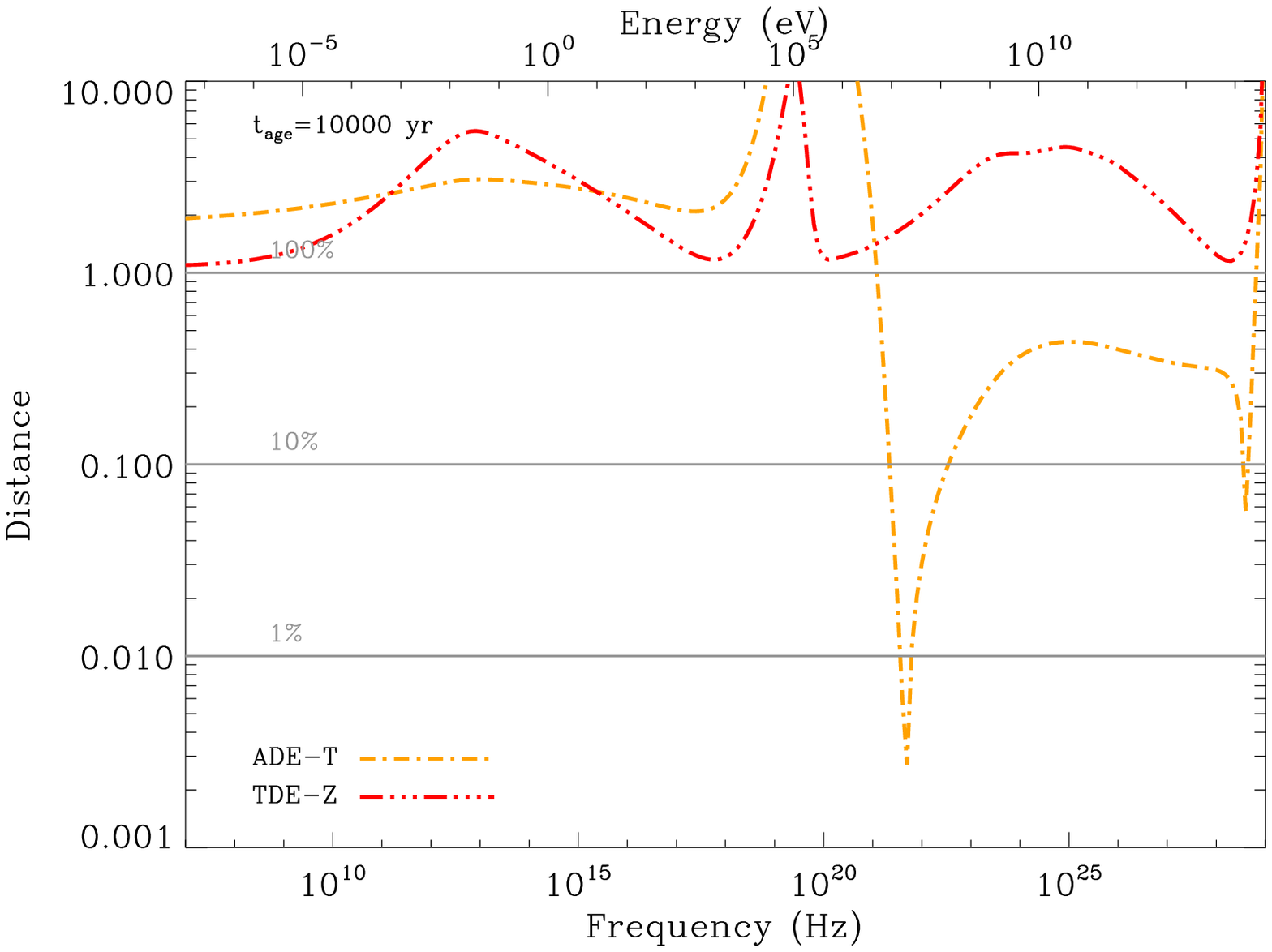}
\end{tabular}
\end{center}
\caption{Relative distance of the results for the photon spectrum between the complete model, the ADE-T, and TDE-Z approximations for the Crab Nebula in different ages.  }
\label{rel-spec}
\end{figure*}

It is clear that at the current age, and particularly due to the fact of the strong SSC domination of the Crab nebula, one can find acceptable sets of parameters in both approximated models that fit the data well. However, this does not mean that the time evolution of these models would be similarly close to the complete analysis.
Figure~\ref{comp-resu-elec} and~\ref{comp-resu-spec} compare the evolution of the results of the complete model and the  approximate  ones, for the electron population and photon spectra, respectively.  Differences increase with the time elapsed off the normalization age (the current Crab nebula), and are clear at a few hundred and a few thousand years. To have a better idea on how the spectra are changing, we use our model as reference data and calculate the relative theoretical distance of the ADE-T and TDE-Z models with respect to it as a function of frequency, both for the electron population and spectrum. We thus compute the theoretical distance as
$
Distance = {|complete - approximate|}/{complete},
$
so that $distance$ 
times 100\% is the percentile value of the deviation.
These results are given in Figure \ref{rel-elec} and \ref{rel-spec}. The dips visible in these figures correspond to the crossing of the curves (lepton Lorentz factors or photon energies where both the approximate and complete models, as plotted in
Figures \ref{comp-resu-elec} and \ref{comp-resu-spec}, coincide). The recovering of the curves after these dips correspond to the use of the absolute value in the {\it Distance} definition. Visually removing these dips gives an idea of the average deviation between the  approximate  and complete models across all energies. Note that the peaks in the relative  distance evolution are broader, and represent bona-fide
increases in the deviation of the approximate results.

Regarding the underlying electron population, we see that differences among models range from 10\% to 100\% and beyond.
The TDE-Z model have a more deviating behavior than theADE-T models at later ages.
Regarding the photon spectrum deviation, we find that 
for ages close to $t_{age}= 940$ yr and lower, the relative distance of all models with respect to the complete one is below 40\% with the exception of the frequency
range between $10^{22}$ and $10^{23}$ Hz, where there is a transition between the synchrotron  and IC dominated radiation. At larger times, deviations can be larger than
100\%. From 2 to 10 kyr, the
relative distance in the optical range and in gamma-rays increases with age. 
As soon as the Crab nebula is let to evolve beyond a few thousand years, and consistently with the results found for the electron population, the relative distance between the spectra of the complete and the  approximate  models goes up to a factor of a few (i.e., percentile distance is a factor of a few 100\%) over large portions of the electromagnetic spectrum. These changes in the nebula evolution are only the by-product of the approximations used in the models and do not represent the expected behavior of the source.

\section{Concluding remarks}

In this work we have introduced 
a leptonic, time-dependent model of PWNe. 
We have considered 
the complete time-energy dependent diffusion-loss equation to compute the lepton population. 
Full Klein-Nishina cross-section for multiple-photon-field inverse Compton, bremsstrahlung, synchrotron,
and self-synchrotron Compton spectra were computed, and their corresponding losses were considered.
The model has allowed, based on fitting against the injection parameters and magnetic field,
to reproduce the current data for the Crab nebula
from radio to TeV. We find that an energy break at $\sim 0.35$ TeV, a high (low) energy index of 2.5 (1.5),
and a magnetic field of 97 $\mu$G (in complete agreement with morphological studies) fits the data perfectly. 
Filamentary structures or flares have not been treated in our model. Other losses, associated with neutrino emission or hadronic
interactions were not taken into account in any of the cases presented.

With the complete model at hand,  we have analyzed which are the consequences of approximations when models ignore losses, photon backgrounds, or escape processes.
In particular, we analyzed the impact of different approximations made at the level of the diffusion-loss equation that allows converting it in advective or in  a time-derivative-only one. Swinging on the parameters one can achieve a relatively good fit to the data of the Crab nebula today. However,  the time-evolution of the electron population and the photon spectrum deviation from the complete analysis is larger than 100\% for these same models, when they evolve in time off the normalization age. This puts in evidence the risks of considering approximations when studying time evolution, as well as,
equivalently, when members of a population observed at different ages are analyzed with the intention of extracting statistical conclusions.

\subsection*{Acknowledgments}

This work was supported by the  grants AYA2009-07391 and SGR2009-811, as well as the Formosa program
TW2010005 and iLINK program 2011-0303.  We acknowledge comments by
W. Bednarek, S. Tanaka, and H. Li.

\section*{Appendix}

\subsection*{Energy losses}

\subsubsection*{Synchrotron losses}

In terms of the Lorentz factor of the lepton, we consider the synchrotron losses described by
\begin{equation}
\label{synclosses}
\dot{\gamma}_{syn}(\gamma,t)=-\frac{4}{3}\frac{\sigma_T}{m_e c}U_B(t)\gamma^2 ,
\end{equation}
where $\sigma_T=(8\pi/3)r^2_0$ is the Thomson cross section, $r_0$ is the electron classical radius, and $U_B(t)=B^2(t)/8\pi$ is the energy density of the magnetic field. We have also assumed that the particles are relativistic, i.e.,  $\beta \simeq 1$.

\subsubsection*{Inverse Compton losses}

We consider the inverse Compton losses described by means of the exact expression of the Klein-Nishina cross section,
\begin{multline}
\label{iclosses}
\dot{\gamma}_{IC}(\gamma)=-\frac{3}{4}\frac{\sigma_T h}{m_e c}\frac{1}{\gamma^2}\int_0^\infty \nu_f \mathrm{d}\nu_f\\
\times \int_0^\infty \frac{n(\nu_i)}{\nu_i} f(q,\Gamma_{\varepsilon}) \theta(1-q) \theta \left(q-\frac{1}{4\gamma^2} \right) \mathrm{d}\nu_i,
\end{multline}
\\
where $h$ is the Planck constant, $\nu_{i,f}$  are the initial and final frequencies of the scattered photons, $\theta$ is the Heaviside step function, and $n$ is
 the photon background distribution.  The other terms are defined as usual, 
\begin{equation}
\label{f}
f(q,\Gamma_{\varepsilon})=2q\ln q+(1+2q)(1-q)+\frac{1}{2}\frac{(\Gamma_{\varepsilon}q)^2}{1+\Gamma_{\varepsilon}q},
\end{equation}
with $
\Gamma_{\varepsilon}={4 \gamma h \nu_i}/{m_e c^2},
$
and $
q={h \nu_f} /({\Gamma_{\varepsilon}(\gamma m_e c^2-h \nu_f)}).
$
The magnitude $\Gamma_{\varepsilon}$ indicates the regime of the IC energy losses. If  $\Gamma_{\varepsilon} \ll 1$, the scatter happens in the Thomson limit and for
$\Gamma_{\varepsilon} \gg 1$, it happens in the extreme Klein-Nishina limit.

\subsubsection*{Bremsstrahlung energy losses}

The general expression for the bremsstrahlung losses is
\begin{equation}
\label{bremseq}
\dot{\gamma}_{Brems}=-N v \int_0^{\gamma-1} k \frac{\mathrm{d}\sigma}{\mathrm{d}k}\mathrm{d}k,
\end{equation}
where $N$ is the number density of particles in the medium, $v$ is the velocity of the electrons, $k=h\nu/mc^2$ is the photon energy in units of the electron rest energy
 and $\mathrm{d}\sigma/\mathrm{d}k$ is the bremsstrahlung differential cross section. The velocity $v$ can be expressed in terms of the Lorentz factor as $ v=c {\sqrt{\gamma^2-1}}/{\gamma}$.
  We consider two contributions in the bremsstrahlung losses: the electron-atom bremsstrahlung due to the interaction of the electron with the electromagnetic field produced
by the ionized nuclei of the interstellar medium (ISM) and the electron-electron bremsstrahlung produced by the electrons also present in the ISM. This second contribution is
the most important and increases with energy, but we include the electron-atom bremsstrahlung for completeness at lower energies.

In the case of the electron-atom bremsstrahlung,
an accurate approximation for the integral above in the whole energy range is 
(Haug 2004)
\begin{multline}
 \int_0^{\gamma-1} k \frac{\mathrm{d}\sigma}{\mathrm{d}k}\mathrm{d}k \simeq \frac{3}{\pi} \alpha \sigma_T Z^2 \frac{\gamma^3}{\gamma^2+p^2} \left[\frac{\gamma}{p} \ln(\gamma+p) -\frac{1}{3} \right. \\
 \left. +\frac{p^2}{\gamma^6} \left(\frac{2}{9} \gamma^2-\frac{19}{675} \gamma p^2-0.06 \frac{p^4}{\gamma} \right) \right],
\end{multline}
where $\alpha \simeq 1/137$ is the fine-structure constant, $Z$ is the atomic number, and $p=\sqrt{\gamma^2-1}$ is the linear momentum of the electron. 
The electron-atom bremsstrahlung energy losses have the form
\begin{multline}
\dot{\gamma}_{Brems}^{e-a}=-\frac{3}{\pi} \alpha \sigma_T c S \frac{\gamma^2}{\gamma^2+p^2} \left [\gamma \ln(\gamma+p)-\frac{p}{3} \right. \\
\left. +\frac{p^3}{\gamma^6} \left(\frac{2}{9}\gamma^2-\frac{19}{675}\gamma p^2-0.06 \frac{p^4}{\gamma} \right) \right],
\end{multline}
with
\begin{equation}
\label{s}
S=\sum_Z Z^2 N_Z=N_H \left[1+\sum_{Z \ge 2} \left(\frac{N_Z}{N_H} \right)Z^2 \right],
\end{equation}
and
where $N_H$ is the number density of hydrogen in the medium and $N_Z$, the number density of the other elements with atomic number $Z$. 

The electron-electron Bremsstrahlung loss is (Blumenthal
 and Gould 1970) 
\begin{equation}
\!
\dot{\gamma}_{Brems}^{e-e}=-\frac{3\alpha}{2\pi} \sigma_T c \!
\left(\sum_Z Z N_Z \right) \! 
\frac{p}{\gamma} (\gamma-1) \!
 \left[\ln(2\gamma)-\frac{1}{3} \right].
\end{equation}

\subsubsection*{Adiabatic energy losses}

The relativistic form of the adiabatic losses is
 \begin{equation}
 \label{adlosses}
\dot{\gamma}_{ad}=-\frac{1}{3} \left(\vec{\nabla} \cdot \vec{v} \right) \gamma.
\end{equation}
We consider the PWN as an uniformly expanding sphere, so the expansion velocity
 of the gas can be written as
\begin{equation}
\label{vel}
v(r)=v_{PWN}(t) \left[\frac{r}{R_{PWN}(t)} \right].
\end{equation}
Applying the divergence operator in spherical coordinates, we get
$
\vec{\nabla} \cdot \vec{v}=({1}/{r^2}) ({\partial v(r)}/{\partial r})
= 3 ({v_{PWN}(t)}/{R_{PWN}(t)}),
$
and substituting in equation~\ref{adlosses}, the adiabatic energy losses are
$
\dot{\gamma}_{ad}(\gamma,t)=- ({v_{PWN}(t)}/{R_{PWN}(t)})\gamma.
$
We use the free expanding expansion of the PWN given by van der Swaluw (2001). The radius of the
 PWN is given by
 \begin{equation}
 \label{rpwn}
 R_{PWN}(t)=C \left(\frac{L_0 t}{E_0} \right)^{1/5} V_0 t,
 \end{equation}
 where $V_0$ is the velocity of the front of the ejecta and has the form
 \begin{equation}
 V_0=\sqrt{\frac{10 E_0}{3 M_{ej}}}.
 \end{equation}
 $E_0$ and $M_{ej}$ are the energy of the supernova explosion and the mass ejected respectively. The constant C is written as
 \begin{equation}
 C=\left(\frac{6}{15(\gamma_{PWN}-1)}+\frac{289}{240} \right)^{-1/5},
 \end{equation}
 with $\gamma_{PWN}=4/3$ since we consider the PWN material as a relativistically hot gas. The velocity of expansion can be easily obtained doing the derivative
 of equation (\ref{rpwn}). Applying the expressions for $R_{PWN}(t)$ and $v_{PWN}(t)$, we get
\begin{equation}
\gamma_{ad}=-\frac{6}{5} \frac{\gamma}{t},
\end{equation}
which differs a factor $6/5$ from a ballistic approximation.
 
The evolution
of the PWN depends also on physical parameters of the previous supernova (SN) event, like the energy of the explosion and the ejected mass. 
The age of the pulsar has to be less than the Sedov time, which can be calculated as (Gaensler \& Slane 2006)
\be
\!\!
\left( \frac{ t_{Sed} }{ {\rm kyr}} \right) \!
 \! \simeq 7  \! \left(\frac{M_{ej}}{10 M_{\odot}} \right)^{5/6}  \!\!
 \left(\frac{E_{SN}}{10^{51}\,{\rm erg}} \right)^{-1/2} \!\!
  \left(\frac{n_0}{{\rm cm}^{-3}} \right)^{-1/3},
\ee
where $M_{ej}$, $E_{SN}$ is the mass and the energy ejected in the supernova explosion, and $n_0$ is the number density of the medium. After the Sedov time the PWN is not expanding freely due to the interaction with the reverse shock of the supernova remnant
(SNR) and a dynamic model is needed to account for its expansion.

\subsection*{Photon spectra}

We briefly summary here the expressions that the code uses to compute the different luminosities.

\subsubsection*{Synchrotron spectrum}
 
 The synchrotron luminosity is 
(Ginzburg \& Syrovatskii 1965, Blumenthal
 and Gould 1970)
\begin{equation}
\label{synclum}
L_{syn}(\nu,t)=\int_{0}^{\infty} N(\gamma,t)P_{syn}(\nu,\gamma,B(t)) \mathrm{d}\gamma,
\end{equation}
where $P_{syn}(\nu,\gamma,B(t))$ is the power emitted by one electron spiraling in a magnetic field
 \begin{equation}
P_{syn}(\nu,\gamma,B(t))=\frac{\sqrt{3}e^3 B(t)}{m_e c^2}F \left(\frac{\nu}{\nu_c(\gamma,B(t))} \right),
\end{equation}
where $\nu_c$ is the critical frequency, $F(x)=x \int_x^\infty K_{5/3}(y) \mathrm{d}y$, and  $K_{5/3}(y)$ is the modified Bessel function of order $5/3$. 
 $N(\gamma,t)$ is calculated solving the diffusion equation explained in Section~\ref{diffusion}. The dependence in time points the need to recall that the magnetic field
 has to be computed at the same age as the luminosity.

\subsubsection*{Inverse Compton spectrum}

The scattered photon spectrum per electron is Blumenthal
 and Gould 1970)
\begin{equation}
P_{IC}(\gamma,\nu,t)=\frac{3}{4} \frac{\sigma_T c h \nu}{\gamma^2} \int_0^\infty \frac{n(\nu_i)}{\nu_i}f(q,\Gamma_{\varepsilon}) \mathrm{d}\nu_i,
\end{equation}
where $\nu_i$ is the initial frequency of the scattered photon, $\nu$ is the final frequency of the photon after scattering, and $n(\nu_i)$ is the photon target field distribution. The expressions for
 $f(q,\Gamma_{\varepsilon})$, $\Gamma_{\varepsilon}$ and $q$ were given above. To obtain the luminosity, we multiply by the electron distribution and integrate in the whole energy range, obtaining 
\begin{equation}
\label{iclum}
L_{IC}(\nu,t)=\frac{3}{4}\sigma_T c h \nu \int_0^\infty \frac{N(\gamma,t)}{\gamma^2} \mathrm{d}\gamma \int_0^\infty \frac{n(\nu_i)}{\nu_i}f(q,\Gamma_{\varepsilon}) \mathrm{d}\nu_i.
\end{equation}

\subsubsection*{Synchrotron Self-Compton spectrum}

We make use of Eq. (\ref{iclum}) and the target photon field density given by (see Atoyan \& Aharonian 1996)
\begin{equation}
\label{nssc}
n_{SSC}(\nu,R_{syn}(t),t)=\frac{L_{syn}(\nu,t)}{4\pi R_{syn}^2(t) c} \frac{\bar{U}}{h \nu}
\end{equation}
where $R_{syn}(t)$ is the radius of the volume where the synchrotron radiation is produced,
and $\bar{U} \simeq 2.24$ is the mean in a spherical volume of the function $U(x)$, given by
\ba
\bar{U}&=&\frac{\int_0^\frac{R_{syn}(t)}{R_{PWN}(t)} x^2 U(x) \mathrm{d}x}{\int_0^\frac{R_{syn}(t)}{R_{PWN}(t)} x^2 \mathrm{d}x} \nonumber \\
&=&3 \frac{R_{PWN}^3(t)}{R_{syn}^3(t)} \int_0^\frac{R_{syn}(t)}{R_{PWN}(t)} x^2 U(x) \mathrm{d}x,
\ea
where 
\begin{equation}
U(x)=\frac{3}{2} \int_0^\frac{R_{syn}(t)}{R_{PWN}(t)} \frac{y}{x} \ln \frac{x+y}{|x-y|} \mathrm{d}y.
\end{equation}
The function U(x) was given by Atoyan \& Nahapetian (1989) to compute the number density of photons at a given distance (in this case, $R_{syn}(t)$) assuming isotropic
emissivity of the synchrotron radiation in a spherical source. The value of this function at $x=0$ is $3$ and decreases until $1.5$ when $x=1$. If we consider a photon
distribution with a radius greater than the radius of the PWN, then $U(x) \simeq 1/x^2$ and $n(\nu,R_{PWN},t)=L(\nu,t)/(4 \pi R_{PWN}^2 c)$. In this case, the photon field
would depend on the PWN radius. For simplicity, we assume $R_{syn}/R_{PWN}=1$ for all cases.

\subsubsection*{Bremsstrahlung spectrum}

The bremsstrahlung luminosity is computed as 
\begin{multline}
\label{bremslum}
L_{Brems}(\nu,t)=\frac{3}{2 \pi} \alpha \sigma_T h c S \int_0^\infty \frac{N(\gamma_i)}{\gamma_i^2} \\
\times \left(\gamma_i^2+\gamma_f^2-\frac{2}{3} \gamma_i \gamma_f \right) \left(\ln \frac{2 \gamma_i \gamma_f m c^2}{h \nu}-\frac{1}{2} \right) \mathrm{d} \gamma_i,
\end{multline}
where $S$ is given in Eq. (\ref{s}), $\gamma_{i,f}$ are the Lorentz factors of the initial and final electron passing through a medium containing different kinds of particles and producing photons with energy $h \nu$.
The kinematic condition $\gamma_i-\gamma_f=h \nu / m_e c^2$ fix the final energy of the electron in the integral given an initial energy $\gamma_i$ and the energy
of the photon produced $h \nu$.

\label{lastpage}
\end{document}